\documentclass[usenatbib]{mnras}
\usepackage{graphicx}
\usepackage[usenames,dvipsnames]{xcolor}
\usepackage{times,verbatim}
\usepackage[fleqn]{amsmath}
\newcommand{\lae}{\lower 2pt \hbox{$\, \buildrel {\scriptstyle <}\over {\scriptstyle
\sim}\,$}}
\newcommand{\gae}{\lower 2pt \hbox{$\, \buildrel {\scriptstyle >}\over {\scriptstyle
\sim}\,$}}

\begin{document}
\title{Swift J1644+57: an Ideal Test Bed of Radiation Mechanisms in a Relativistic Super-Eddington Jet}
\author[P. Crumley, W. Lu, R. Santana, R. A. Hern{\'a}ndez, P. Kumar,  S. Markoff]{P. Crumley,$^{1,2,3}$\thanks{E-mail:
    p.k.crumley@uva.nl, pk@surya.as.utexas.edu}
  W. Lu,$^2$ R. Santana,$^2$
  R. A. Hern{\'a}ndez,$^{1,2}$, P. Kumar$^2$\footnotemark[1], S. Markoff$^3$ \\ 
  $^1$Physics Department, University
  of Texas at Austin, Austin, TX 78712\\ $^2$Astronomy Department,
  University of Texas at Austin, Austin, TX 78712\\ $^3$Anton Pannekoek Institute for Astronomy, University of Amsterdam, P.O. Box 94249, 1090 GE Amsterdam, the Netherlands}

\maketitle
\begin{abstract} 
Within the first 10 days after Swift discovered the jetted tidal disruption event (TDE) Sw J1644+57, simultaneous observations in the radio, near-infrared, optical, X-ray and gamma-ray bands were carried out. These multiwavelength data provide a unique opportunity to constrain the emission mechanism and make-up of a relativistic super-Eddington jet. We consider an exhaustive variety of radiation mechanisms for the generation of X-rays in this TDE, and rule out many processes such as SSC, photospheric and proton synchrotron. The infrared to gamma-ray data for Sw J1644+57 are consistent with synchrotron and external-inverse-Compton (EIC) processes provided that electrons in the jet are continuously accelerated on a time scale shorter than \(\sim 1\%\) of the dynamical time to maintain a power-law distribution. The requirement of continuous electron acceleration points to magnetic reconnection in a Poynting flux dominated jet. The EIC process may require fine tuning to explain the observed temporal decay of the X-ray lightcurve, whereas the synchrotron process in a magnetic jet needs no fine tuning for this TDE.

\end{abstract}
\section{Introduction}
Narrowly collimated, relativistic jets are responsible for some of the most luminous and energetic astrophysical high-energy sources: gamma-ray bursts (GRB) and active galactic nuclei (AGN). How the jet's energy is converted into the particles' randomized kinetic energy, and how the particles convert their kinetic energy into the observed non-thermal X-rays and $\gamma$-rays are unanswered questions in astrophysics.\footnote{For a recent paper that deals with this same question, but with a very different approach and technique, see \citet{2015MNRAS.450..183S}.} The answers to these questions depend on the nature of jet. If the jet's energy is carried primarily by the baryons, shocks can dissipate the jet's energy and heat the particles \citep[e.g.,][]{1994ApJ...430L..93R}. If the jet is magnetically dominated at the region where it radiates, the ability for shocks to dissipate the jet's energy is diminished. In Poynting dominated jets, magnetic reconnection is a more likely way to produce the non-thermal particles. \citep[e.g.,][]{2002A&A...391.1141D} On March 28, 2011 the Swift Burst Alert Telescope (BAT) triggered on Swift J164449.3+573451, hereafter Sw J1644+57, a new type of astrophysical X-ray transient---a relativistic tidal disruption event. The excellent data of Sw J1644+57 afford a new opportunity to examine how particles are accelerated and radiate away their energy in relativistic jets.

When a star wanders too close to the super-massive black hole (SMBH) at its galactic center, the tidal forces from the black hole grow rapidly. At some point, the star's self-gravity is no longer capable of resisting the tidal forces of the black hole. If a star passes by a SMBH within this tidal radius, the star will be torn apart and swallowed by the SMBH in a tidal disruption event (TDE). In order to conserve angular momentum, roughly half of the star's mass remains gravitationally bound to the black hole, while the rest is flung away. The bound gas will return to pericenter and form an accretion disk \citep{1982ApJ...262..120L}. The mass falls back onto this accretion disk at a super-Eddington accretion rate. The accretion rate decays in a  characteristic manner, as $t^{-5/3}$ \citep{1988Natur.333..523R, 1989IAUS..136..543P}. In some TDEs, a relativistic jet is launched, re-igniting a dormant quasar.

While originally a theoretical curiosity, in the last 10 years or so there have been many claims of an observed tidal disruption event\footnote{\nocite{website:tde-claims} See J. Guillochon's http://astrocrash.net/resources/tde-catalogue/ for a catalog of astrophysical transients claimed to be a TDE.}.  In the two TDEs discovered by the Swift BAT, Sw J1644+57 and Sw J2058.4+0516, the TDE launched a relativistic, super-Eddington jet that was observed along the jet axis  \citep{Bloom2011, Burrows2011,2012ApJ...753...77C}. These objects are the gold standard for TDEs---the only known super-Eddington, relativistic TDEs where we were able to see the early time X-ray emission from the jet. We focus on Sw J1644+57 because it occurred at a smaller redshift ($z=0.354$) than J2058 ($z=1.185$). The two events released a similar amount of energy; however the observed flux is much brighter for J1644 because it is $\sim4$ times closer than J2058. Because Sw J1644+57 is much closer, the quality of its data is much better.

Sw J1644+57 was luminous ($L_X \sim 10^{48}\ \mathrm{erg/s}$), highly variable ($\delta t\sim 100$ s), long-lasting (around 500 days), and emitted most of its energy in X-rays and soft $\gamma$-rays in a hard, non-thermal power-law ($f_\nu\propto\nu^{-0.8}$) \citep{Burrows2011,Bloom2011}. Detection of a coincident source in the optical, infrared, and radio wavelengths revealed that Sw 1644+57 was an extra-galactic source, located within the nucleus of its host galaxy \citep{Levan2011,Zauderer2011}. Archival images showed that the host galaxy was quiescent before Swift triggered on Sw J1644+57. Shortly following the trigger, the mass of the black hole that caused the tidal disruption was estimated using rough variability time arguments as being $10^6$ to $10^7\ M_\odot$ \citep{Burrows2011, Levan2011}. Now that the optical flux from the TDE lies below the flux of the host galaxy, it is possible to estimate the mass of the black hole using the black hole mass - bulge relationship, which yields a result of $M_\bullet = 10^{6.7\pm0.4}M_\odot$, consistent with the other estimates \citep{2015ApJ...808...96Y}.

Swift continued to observe variable X-ray emission from Sw J1644+57 for $\sim 500$ days after the trigger, and the X-ray flux from Sw 1644+57 decreased in a manner consistent with $t^{-5/3}$. These prompt X-rays are thought to be produced in a relativistic, $\Gamma\gae10$, jet. After $\sim 500$ days, there was a steep drop in the X-ray flux. This drop off has been widely interpreted as the jet producing the prompt X-rays turning off. All of the evidence points to Sw J1644+57 being the first observed tidal disruption of a stellar mass star by a super-massive black hole with a relativistic jet pointed towards us. 

The quality and breadth of Sw J1644+57's broadband data make it an ideal astrophysical system to test different emission mechanisms. The long-lasting prompt emission allowed for a wealth of simultaneous data, with multiple instruments taking measurements in the radio, infrared, optical, X-rays, and soft $\gamma$-rays, as well as placing tight upper limits in the high-energy $\gamma$-rays. The radio and X-ray data of J1644+57 come from different sources; the radio data is well-modeled as being produced in the interaction between the relativistic jet producing the X-rays with the external medium surrounding the SMBH \citep{2012MNRAS.420.3528M, Berger2012,Zauderer2013}. The radio light curve is flat for $\sim100$ days, a fact that causes the energy in the forward shock to appear to increase by a factor of $\sim 20$ \citep{2013ApJ...770..146B}. It is difficult to explain this as late time activity in the jet engine, because the X-ray flux continues to decline as $t^{-5/3}$, belying the idea that a significant amount of energy is being injected into the system. Increasing the forward shock's energy by having slower moving ejecta catch up to the decelerating forward shock requires a large amount of energy. To solve this issue, some alternate models have been suggested. \citet{2013MNRAS.434.3078K} suggested that the radio-producing electrons are inverse Compton cooled by the observed X-ray photons. The rate of cooling decreases as the X-ray luminosity decreases, producing a flat radio light curve. \citet{2015arXiv150100361M} suggested that the flat light curve is produced by spine-sheath jet structure in Sw J1644+57. When considering inverse Compton cooling and adding up the contribution of the forward shock from a faster ($\Gamma\sim10$), narrower, core jet and a slower ($\Gamma\sim2$), poorly collimated outflow, the observed radio light curve can be reproduced. Regardless of the details, the general picture is clear: the observed radio data is produced by the interaction between the relativistic jet responsible for the X-ray emission and the external medium. The radio data of Sw J1644+57 is analogous to $\gamma$-ray burst afterglows, and as in $\gamma$-ray bursts, modeling the prompt emission of Sw J1644+57 is fraught with difficulty.  

Although the temporal behavior of the prompt X-rays in J1644+57 is explained by the rate that the tidally disrupted stellar gas returns to an accretion disk, how these non-thermal X-rays are produced is not understood. Any model of the prompt X-rays will face several challenges. As we show in \S \ref{sec:Synch} and \ref{sec:IC}, the electrons are strongly cooled by both synchrotron and inverse Compton radiation, and may cause excess flux at lower-energy bands unless self-absorbed. Additionally, the upper limits provided by the Fermi Large Area Telescope (100 MeV-10 GeV) and VERITAS at $\gae$500 GeV require that the self-Compton high energy $\gamma$-rays be suppressed either through $\gamma+\gamma$ pair production or Klein-Nishina effects. These requirements limit the possible radii where the X-rays can be emitted by the jet. A strict lower limit to the emission region is the Schwarzschild radius of the SMBH, $3\times 10^{11} M_{\bullet,6} \mathrm{cm}$, but variability time arguments would place the radius of emission closer to $\sim 4\times10^{14} \Gamma_1^2\delta t_2\ \mathrm{cm}$. Sw J1644+57 also has a tight energy budget, since we know the total mass that accreted onto the black hole cannot exceed the mass of the star that was disrupted, $\sim 1 M_\odot$. Therefore production of the X-rays must be energy efficient; otherwise the total energy required exceeds the energy budget of the TDE. However, these challenges are really opportunities; they allow us to robustly rule out and constrain emission models in ways not possible in other astrophysical systems. 

Previous works have modeled the emission of Sw J1644+57. Attempts were made to model X-ray emission with models commonly used in active galactic nuclei and $\gamma$-ray bursts: external inverse Compton and synchrotron from  internal shocks \citep{2015ApJ...798...13L, 2012MNRAS.421..908W}. These models are focused on the X-ray and soft $\gamma$-ray emission, and they ignore simultaneous measurements made in other wavelengths. Broadband fitting of the entire spectral energy distribution was done by \citet{Bloom2011}, who used an different external Compton model and fitted the spectrum at early times but not during the brightest flares of Sw J1644+57. \citet{Burrows2011} modeled the entire spectral energy distribution during periods of flaring, average and quiescent activity at early times. They concluded that the most probable explanation for the X-rays is a Poynting-dominated jet where the electrons are accelerated continuously in magnetic reconnection regions. 

The goal of this paper is to determine the requirements on different emission models to match the observed properties of Sw J1644+57. We use the broadband data Sw J1644+57 to critically assess the viability of as many models of producing non-thermal emission in relativistic jets as possible. By fitting the non-thermal X-rays of Sw J1644+57 we hope to learn something about the jet that produced those X-rays. We first we limit ourselves to situations where the electrons are accelerated impulsively i.e., any scenario where the electrons are quickly accelerated in a small region of the source. After being accelerated, the electrons then leave the acceleration region and cool either radiatively, via synchrotron and inverse Compton radiation, or adiabatically. While shocks are typically invoked as a way to accelerate electrons impulsively, our arguments apply to any source where the electrons radiate most of their energy after they are finished being accelerated, and they are not re-accelerated after the initial acceleration. In the context of impulsive acceleration we consider the electron synchrotron process (\S \ref{sec:Synch}), the proton synchrotron process (\S \ref{sec:Pro_synch_TDE}),and internal inverse Compton radiation (\S \ref{sec:IC}). Internal inverse Compton radiation is where the electrons which up-scatter the seed photons are not necessarily the same population of  electrons that produce the seed field, but the seed field is produced inside of the jet.\footnote{Synchrotron self-Compton (SSC) is a particular example of the more general "internal" inverse Compton process considered here.} Besides these models, we also consider external inverse Compton scattering of optical and UV photons produced in an optically thick wind coming off the disk (\S \ref{sec:EIC}), and photospheric radiation (\S \ref{sec:Photospheric}), where a thermal spectrum is reprocessed by hot electrons in the jet. Finally, we apply the the magnetic-reconnection model developed in \citet{KumarCrumley2015} to Sw J1644+57 (\S \ref{sec:Poynting}). 
\section{Overview of Sw J1644+57 properties}
The prompt emission of J1644+57 lasted for around 500 days, and had a total fluence in XRT band for the first $\sim50$ days of $7.1\times10^{-4}\ {\rm erg\ cm^{-2}}$, which corresponds to a isotropic energy release of $2\times10^{53}\ \mathrm{erg}$. During the first ten days the TDE experienced a period of intense X-ray flaring, with each individual flare lasting $\sim10^3$ seconds, with peak luminosities greater than $10^{48}\ \mathrm{erg/s}$ \citep{Burrows2011}. After this time, the X-ray light curve shows a decline consistent with $t^{-5/3}$ lasting $\sim$ 500 days. After 500 days, the X-ray flux shows a precipitous drop. This steep drop off is interpreted to be the jet turning off. 

During the early intense flaring, the Swift data shows an absorbed single power law in both the XRT and BAT band, with a spectral index  $\beta$ of $\sim0.8$, \(f_\mathrm{\nu}\propto\nu^{-0.8}\), and an unabsorbed integrated flux from 1-10 keV of $6.5\times 10^{-10}\ \mathrm{erg\ cm^{-2}\ s^{-1}}$ \citep{Burrows2011}. This integrated flux corresponds to a specific flux $f_\nu$ of 0.1 mJy at 1 keV and an isotropic luminosity of $L_X=2.6\times10^{47} \mathrm{erg\ s^{-1}}$. \citet{Saxton2012} measured the late time properties of the XRT spectrum and they found that $\beta=0.77\pm0.05$ for the first 55 days, consistent with \citet{Burrows2011}. After the first 55 days, the spectrum hardens to $\beta = 0.4$ for the next 100 days, with an average integrated flux in the XRT band $\sim10^{-11} \ \mathrm{erg\ cm^{-2}\ s}$, corresponding to specific flux at 1 keV of $\sim 8\times10^{-4} \ \mathrm{mJy}$. There is some uncertainty in the normalization and spectral index in the X-ray flux, but our findings are not particularly sensitive to the exact values given above. Wherever possible, we try to give analytic estimates of quantities. Therefore, so it is possible to see how our findings depend on our fiducial values for Sw J1644+57. The peak luminosity of Sw J1644+57 was about $3\times10^{48}\ \mathrm{erg\ s^{-1}}$ and the minimum luminosity in the first 10 days was $\sim 10^{46}\ \mathrm{erg \ s^{-1}}$. The total isotropic energy release in X-rays was $2\times10^{53}$ erg. Considering a solar mass progenitor, the radiation mechanism must be $\gae1$\% efficient; otherwise the energy requirements will exceed the energy budget. 

Since Sw J1644+57 is located in the nucleus of its host galaxy, the optical extinction is large, $A_v\approx 4.5$ \citep{Burrows2011}. We therefore restrict ourselves to the K-band measurements where the extinction problem should be minimized. For an $A_v\approx 4.5$, we expect that the dereddened K-band values to be $\sim60$\% larger than the observed ones  \citep{2011ApJ...737..103S}. The K-band flux is $f_\nu\sim 0.1 \ \mathrm{mJy}$ 2 days after trigger. The optical flux decreases steadily for the first ten days, rises to a peak at $\sim50$ days, and then decays again \citep{2015arXiv150908945L}. The optical measurements also allow for the redshift to be measured, $z=0.354$. This redshift corresponds to a luminosity distance, $d_L=5.8\times10^{27}\ \mathrm{cm}$ in a flat $\Lambda$CDM cosmology with $\Omega_\Lambda=0.714$, $\Omega_M = 0.286$, and $H_0=69.6$.

In the hard $\gamma$-rays, A non-detection by Veritas sets an upper limit on the $\gamma$-ray flux at 500 GeV of $1.4 \times10^{-12}\ \mathrm{erg\ cm^{-2}\ s^{-1}}$ \citep{2011ApJ...738L..30A}. The Fermi LAT finds a similar upper limit on the $\gamma$-ray flux between 100 MeV-10 GeV, $2.7\times10^{-11} \ \mathrm{erg\ cm^{-2}\ s^{-1}}$, a factor of 24 times lower than the observed XRT X-ray flux \citep{Burrows2011}.

The observed radio photons from Sw J1644+57 come from a different process than the X-ray photons. Therefore, we do not try to model the radio data in this paper. However, the radio data can be used to constrain some properties of the jet such as the initial bulk Lorentz factor which is $\sim 10-20$ \citep{2012MNRAS.420.3528M}.

When modeling Sw J1644+57, we simply focus on reproducing the observed XRT-BAT X-ray emission, we do not require that the K-band flux be produced by the same mechanism that produces the X-rays; for instance, the K-band flux could be produced by the accretion disk. Instead we model the X-ray emission and then simply check to see if the K-band flux or $\gamma$-ray upper limits is significantly over produced for a particular set of parameters in a particular model. If that is the case, then the model doesn't work for those parameters. If we have exhausted the entire possible parameter space without finding a solution consistent with the multi-wavelength data of Sw J1644+57, then we have ruled out that model. 
\section{Synchrotron Model of TDE Sw J1644+57}\label{sec:Synch}
In this section, we provide analytical arguments to show the difficulties an isotropic synchrotron model has in reproducing the X-ray emission from Sw J1644+57. Then we describe a new synchrotron self-Compton numerical method, we use to search the available parameter space to see if we can reproduce the observations. This new method self-consistently calculates the inverse Compton and synchrotron emission, accounting for Klein-Nishina effects and also the non power-law electron distribution that may occur below the synchrotron self-absorption Lorentz factor \citep[cf][]{Ghisellini88,deKool1989}. The assumptions we make for the geometry of our jet in our synchrotron model for Sw J1644+57 are similar to the model developed for GRBs presented in \citet{2008MNRAS.384...33K}. We assume it is isotropic and that electrons are only accelerated one time in the jet.

The synchrotron flux at a frequency $\nu$ from a spherically symmetric relativistic source depends on 6 free parameters: $\Gamma$, the bulk Lorentz factor; $B'$, the co-moving magnetic field strength; $N_e$, the number of electrons radiating at $\nu$; $R$, the radius of emission; $\gamma_i$, the typical electron Lorentz factor; and $p$, the electron spectral index.

The synchrotron frequency and peak flux are:
\begin{equation}\label{eq:nu_i}
\nu_i=\frac{qB' \gamma_i^2\Gamma}{2\pi m_e c(1+z)}\approx (1.2\times10^{-8}\ \mathrm{eV})B'\gamma_i^2\Gamma(1+z)^{-1}
\end{equation}
\begin{align}\label{eq:f_i}
f_i&=\frac{\sqrt{3}q^3B'N_e\Gamma(1+z)}{4\pi d_L^2m_ec^2}\\
&\approx (1.8\times10^2\ \mathrm{mJy})N_{e,55}B'\Gamma(1+z)/d_{L,28}^2
\end{align}
where $d_{L}$ is the luminosity distance to the source.  Throughout the paper frequencies $\nu$ are measured in eV, fluxes are measured in mJy, and the convention $(x_n\equiv x/(10^n\ \mathrm{cgs}))$ is used. 

The bulk Lorentz factor $\Gamma$ can be constrained by the variability time, $\delta t$. The conservative upper limit on the variability time of Sw J1644+57 is $\sim100$ seconds \citep{Burrows2011}. Assuming isotropic emission in the rest frame of the jet, the dynamical time must be equal to or smaller than the variability time, \(\delta t \gae t_\mathrm{dyn}=(1+z)R/(2c\Gamma^2)\). Since there is good evidence that the X-ray source is at least mildly relativistic \citep{Bloom2011,Levan2011}, we also require that $\Gamma>2$.
\begin{equation}\label{eq:Gamma}
\Gamma = \max{(4.1(1+z)^{1/2}\delta t_2^{-1/2} R_{14}^{1/2},2)}
\end{equation}
To simplify the expressions, we restrict ourselves to the regime where $R$ is sufficiently large to ensure $4.1(1+z)^{1/2}\delta t_2^{-1/2} R_{14}^{1/2}$ is greater than 2 in this section.

To eliminate $N_{e,55}$ from equation \eqref{eq:f_i}, we define a new quantity $\widetilde{Y}$,
\begin{equation}\label{eq:ComptonY}
\widetilde{Y}\equiv \gamma_i^2\tau_e,
\end{equation}
where $\tau_e$ is the optical depth of the source, 
\begin{equation}\label{eq:tau_i}
\tau_e=\frac{\sigma_TN_e}{4\pi R^2}\approx 53 N_{e,55}R_{14}^{-2}.
\end{equation}
$\sigma_T$ is the Compton cross section. $\widetilde{Y}$ is closely related to Compton $Y$. Compton $Y$ is the ratio of power radiated via the inverse Compton process to the synchrotron process. In the Thomson regime, $Y\equiv \frac{U'_\gamma}{U'_B}$, \textit{i.e.}, the ratio of the energy density in the photons to the energy density of the magnetic field. In the Thomson regime, $\widetilde{Y}$ is smaller than $Y$ by a factor of order unity that depends on the electron index $p$. This small difference is due to the fact that there are electrons with Lorentz factors greater than $\gamma_i$. However, $\widetilde{Y}$ will be greater than Compton $Y$ when $\gamma_i$ is large enough to cause the inverse Compton scattering to be Klein-Nishina suppressed.

The typical Lorentz factor of electrons radiating a flux $f_{\nu_i}$ at $\nu_i$ is calculated using equations (\ref{eq:nu_i}), (\ref{eq:f_i}), and (\ref{eq:Gamma})
\begin{equation}\label{eq:gamma_i}
\gamma_i\approx 
740\ f_{i,\mathrm{mJy}}^{-1/4}
\nu_{i,3}^{1/4}\widetilde{Y}^{1/4}R_{14}^{1/2}
(1+z)^{1/2}d_{L,28}^{-1/2}
\end{equation}
where $\nu_{i,3}$ is the frequency in keV.

The magnetic field required for the electrons with the $\gamma_i$ given by eq (\ref{eq:gamma_i}) to radiate at $\nu_i$
\begin{equation}\label{eq:MagField}
B'\approx (3\times10^{4}\ \mathrm{G})\delta t_2^{1/2}\nu_{i,3}^{1/2}f_{i,\mathrm{mJy}}^{1/2}\widetilde{Y}^{-1/2}R_{14}^{-3/2}\frac{d_{L,28}}{\sqrt{1+z}}
\end{equation}
While (\ref{eq:f_i}) and (\ref{eq:MagField}) gives the necessary number of electrons:
\begin{equation}\label{eq:electronNum}
N_{e,55}\approx 4\times10^{-8}\ \nu_{i,3}^{-1/2}f_{i,\mathrm{mJy}}^{1/2}
\widetilde{Y}^{1/2}R_{14}(1+z)^{-1}d_{L,28}
\end{equation}
The optical depth of the electrons producing the observed XRT radiation is found by substituting eq (\ref{eq:electronNum}) into eq. (\ref{eq:tau_i})
\begin{equation}\label{eq:opt_depth}
\tau_e\approx 1.4\times10^{-6}\ \nu_{i,3}^{-1/2}f_{i,\mathrm{mJy}}^{1/2}
\widetilde{Y}^{1/2}R_{14}^{-1}(1+z)^{-1}d_{L,28}
\end{equation}

Compare the optical depth of the electrons radiating in the XRT band to the optical depth expected if the jet is baryonic, $\tau_{\rm jet}$
\begin{equation}\label{eq:tau_jet}
\tau_\mathrm{jet}\approx \frac{\sigma_TL_j(R/\Gamma)}{4\pi R^2 m_p c^3\Gamma^2}
\approx 1.2\times10^{-2}\ \frac{L_{j,48}}{R_{14}\Gamma_1^3}
\end{equation}

In Sw J1644+57 the flux at 1 keV during the first 10 days fluctuated between $2\times10^{-3}$ and $1$ mJy. The isotropic X-ray luminosity varied between $10^{46}$ and $3\times 10^{48}$ erg/s. The TDE occurred at redshift $z=0.35$ corresponding to a distance $d_L=5.8\times10^{27}$ cm. If we take $L_{j,48} \sim 3$ and $f_{i,-3}\sim 1$ mJy there is a discrepancy between the optical depth of the XRT producing electrons and optical depth expected if the jet is baryonic of a factor  $\sim10^5$ when $\widetilde{Y}\lae10$. Equation (\ref{eq:opt_depth}) implies  $\tau_e \sim 6 \times10^{-7} \widetilde{Y}^{1/2}$ whereas eq (\ref{eq:tau_jet}) implies $\tau_\mathrm{jet}\sim 4\times10^{-2}\Gamma_i^{-3}$. This discrepancy suggests only 1 in $10^5$ electrons are radiating X-rays if $\Gamma\approx 10$. Making the two optical depths equal requires $\Gamma\sim300$, too large for the Sw J1644+57. Alternatively, $\widetilde{Y}$ could be large, $\sim4\times10^4$ with a more modest $\Gamma \sim 50$. While $\widetilde{Y}\sim4\times10^4$ seems too large, remember that $\widetilde{Y}$ could be much larger than the true Compton $Y$ due to the fact $\widetilde{Y}$ does not account for Klein-Nishina suppression. Another way to account for the large difference between eq (\ref{eq:tau_jet}) and  (\ref{eq:opt_depth}) could be that $\gamma_i \gg \gamma_c$, where $\gamma_c$ is the cooling Lorentz factor. We will now show that for the majority of the available parameter space of Sw J1644+57, $\gamma_i\gg \gamma_c$.

Assuming that $Y \lae 1$ so that synchrotron cooling dominates, we find a cooling Lorentz factor of
\begin{equation}
\gamma_c\approx \frac{6\pi m_ec}{\sigma_TB'^2t_\mathrm{dyn}'};
\qquad  t_\mathrm{dyn}'\sim \frac{R}{2c\Gamma}
\end{equation}
Therefore,
\begin{equation}
\gamma_c\approx\frac{12\pi m_e c^2\Gamma}{\sigma_T B'^2R}
\approx\frac{4.6\times10^5\Gamma}{B'^2R_{14}}
\end{equation}
Substituting eq (\ref{eq:MagField}) for $B'$ and (\ref{eq:Gamma}) for $\Gamma$ we find
\begin{equation}\label{eq:gamma_cool}
\gamma_c\approx 2\times 10^{-3}\ \nu_{i,-3}^{-1}f_{i,-3}^{-1}\widetilde{Y}
R_{14}^{5/2}(1+z)^{3/2}\delta t_2^{-3/2}d_{L,28}^{-2}
\end{equation}
Of course $\gamma_c<1$ is not physical; what it means is that electrons are deep in the fast cooling regime, i.e. $\nu_c\ll$1 keV. Being in the fast cooling regime is a problem for Sw J1644+57 for two reasons. First, the observed spectrum $f_\nu\propto\nu^{-0.8}$ at 1 keV is more naturally explained if electrons producing the X-ray photons are uncooled.  If $\nu_c$ is much greater than a keV, then $\beta$ is related to the electron index $p$ by $\beta=(p-1)/2$ and $p$ is be 2.6. If $\nu_c$ is less than a keV then $\beta = p/2$  and $p$ must be 1.6. If $\nu_c$ is still less than a keV after the first 50 days, $p$ must be 0.8. A $p<2$ is unlikely to occur from traditional shock acceleration \citep[cf.][]{1987PhR...154....1B, 2001MNRAS.328..393A, 2008ApJ...682L...5S, 2011ApJ...726...75S}. Second, if $\nu_c$ is much less than 1 keV, the solution runs the risk of causing the predicted K-band flux to be much larger than observed K-Band flux. 

The average flux at 1 keV during the first ten days of Sw J1644+57 was $\sim 0.1$ mJy, while the maximum dereddened K-band flux during the same time was $\sim0.15$ mJy \citep{Levan2011}. Preventing the X-ray electrons from overproducing in the K-band requires either $\gamma_c\sim\gamma_i$, or for the self-absorption frequency, $\nu_a$, to be larger than 0.57 eV (the energy of the K-band photons). It is clear from eq (\ref{eq:gamma_cool}) that $\gamma_c\sim\gamma_i$ is very unlikely, except for extreme parameters for Sw J1644+57. Therefore, the only way to save the XRT producing electrons from cooling and then overproducing the K-band flux is to have $\nu_a>0.57$ eV. We now show it may be possible to avoid overproduction in the K-band through self-absorption by estimating the maximum possible $\nu_a$.

\begin{figure*}
\begin{center}
\includegraphics[width=\textwidth]{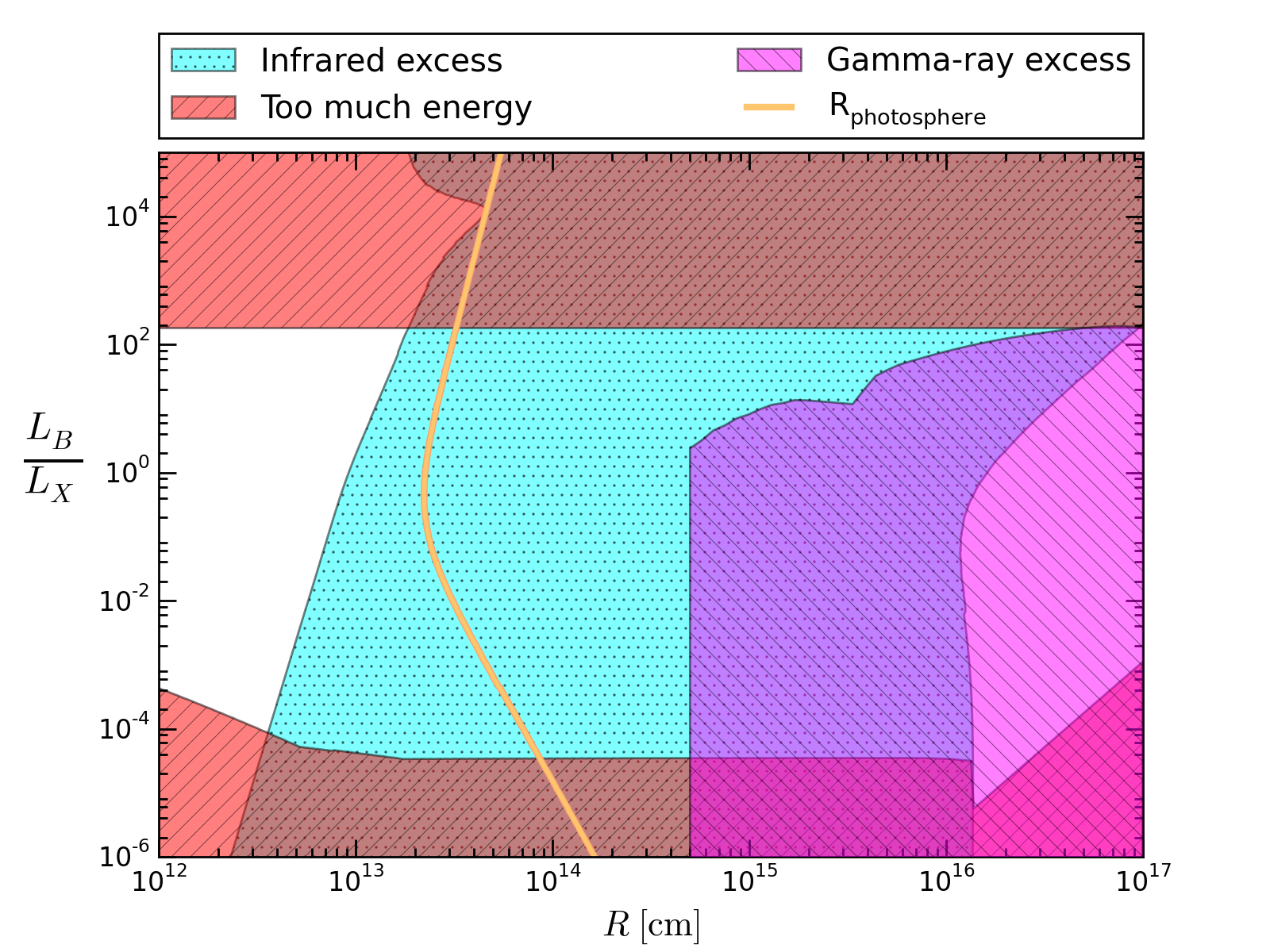}
\end{center}

\caption[Parameter Space available to the Synchrotron-in-Shock model of Sw J1644+57]{This figure shows the various constraints the observations of Sw J1644+57 impose on the synchrotron  model. The $y$-axis is $L_B/L_x$, the luminosity carried by the magnetic field divided by the luminosity in the X-rays. The $x$-axis is the emission radius, $R$. Two regions on the plot require too much energy. In the upper region, the magnetic energy required is too large, and in the lower region, the electrons put too much of their energy into inverse Compton radiation rather than synchrotron, i.e. for the luminosity to be greater than $10^{50}$ erg/s Compton $Y \gae 400$. Note that this occurs at a $L_B/L_x$ smaller than $\sim 2\times10^{-3}$ because of Klein-Nishina effects. LAT excess is where the XRT-BAT X-ray photons would be up-scattered into the LAT band and violate the strict upper limits measured in Sw J1644+57. Infrared excess is the region where cooled electrons will radiate in the K-band, overproducing the observed flux. Overlapping regions are where the synchrotron model will not work for several reasons. The yellow line marks the location of the photosphere. There is one region, deep beneath the photosphere, where the synchrotron model is allowed $R\lae2\times10^{13}\ \mathrm{cm}$. In this region, the synchrotron self-absorption break prevents the K-band flux from being too large.
}
\label{fig:Synch}
\end{figure*}

The maximum possible $\nu_a$ in the synchrotron model will occur when the XRT spectrum can be extrapolated all the way to $\nu_a$ (i.e. $\nu_m\leq\nu_a$). When this is the case, the electrons that dominate the self-absorption have Lorentz factors $\gamma_a$ 
\begin{equation}
\gamma_a \equiv 9100 \nu_a^{1/2} B'^{-1/2}\Gamma^{-1/2} (1+z)^{1/2} \quad \mathrm{if}\quad \gamma_\mathrm{min}<\gamma_a
\end{equation}
The self-absorption frequency in the co-moving frame can be estimated by equating the synchrotron flux at $\nu_a$, $f_{\nu_a}$ to the black-body flux at $\nu_a$ in the Rayleigh-Jeans limit.
\begin{equation}\label{eq:self_absorp_start}
\frac{2\pi \nu_a'^{2}}{c^2}m_e c^2\gamma_a = 
\frac{f_{\nu_a}}{\Gamma}\frac{d_L^2}{(1+z)R^2}
\end{equation}
Note that in eq (\ref{eq:self_absorp_start}) all quantities are in CGS units, but as before primes refers to the quantities measured in the co-moving rest frame. Assuming $f_{\nu_a}=f_i(\nu_a/\nu_i)^{-\beta}$, eliminating $B'$ by using $\widetilde{Y}$, and converting all frequencies to eV measured in the observer frame we find 
\begin{equation}\label{eq:self_absorp_obs_no_mag}
\nu_{a,\mathrm{eV}}^{(5+2\beta)/2}
=1.1\times 10^{-2+3\beta}f^{5/4}_{i,\mathrm{mJy}} \nu_{i,3}^{(4\beta+1)/4}
\frac{\Gamma d_{L,28}^{5/2}R_{14}^{-3}}{\widetilde{Y}^{1/4}(1+z)^{-7/2}}
\end{equation}
When $\beta = 0.8$, the self-absorption frequency depends weakly on all of the free parameters except for $R$. With $f_{i,\mathrm{mJy}}=0.1$, $\nu_i=1$ keV, $z=0.35$, and $d_{L,28}=0.58$, the self-absorption frequency is \begin{equation}
\nu_{a}\approx 1\ \mathrm{eV}\ \widetilde{Y}^{-0.08}\Gamma_1^{0.3}R_{14}^{.9}
\end{equation}
At $R\sim 3\times10^{13}$ cm, $\nu_a\sim 3$ eV, and the self-absorption frequency is large enough to prevent the K-band flux from being too large. Given the above self-absorption frequency, the flux at 0.57 eV is roughly equal to the observed value of $0.15$ mJy, 

\begin{equation}
f_\mathrm{K-band} =  f_{i,\mathrm{mJy}}
(\nu_a/\nu_i)^{-0.8} (0.57/\nu_a)^{5/2} 
\end{equation}
$f_\mathrm{K-band} \approx  0.15 \  \mathrm{mJy}$
when $\nu_a\sim 3$, $\nu_i = 1\ \mathrm{keV}$, $f_i=0.1\ \mathrm{mJy}$.

So while $p<2$ may make it seem somewhat unnatural for the electrons producing the X-ray photons to be cooled, $\nu_c<1\ \mathrm{keV}$ may not cause an overproduction of the K-band flux. Therefore we cannot automatically rule out the synchrotron model of Sw J1644+57; however the analytical calculations provided above argue that any solution would lie in an extreme part of the parameter space. To approach the problem, the self-absorption frequency must be determined very accurately, due to the sensitivity of the K-Band flux to the self-absorption frequency, as well as issues with calculating the synchrotron cooling below the self-absorption frequency. To this end we developed a new numerical code that uses 1-D radiative transfer to calculate the synchrotron and the synchrotron self-Compton spectrum with the main goal of self-consistently and accurately determining the self-absorption frequency and synchrotron flux in the K-band. Additionally, we will calculate the Inverse Compton flux  at LAT and VERITAS energies. The goal of the code is to see if matching the observed synchrotron flux and spectrum at 1 keV causes too much flux in a different frequency band, in violation of observations and upper limits at these frequencies.

We use a methodology that calculates the observed inverse Compton flux and inverse Compton cooling accurately in both the Thomson and Klein-Nishina regime. It numerically integrates over the electron distribution to calculate the synchrotron and inverse Compton flux. The code is capable of calculating the spectrum from a non-power-law electron distribution and will iterate until the predicted emission is self-consistent with the cooling rate predicted from the spectrum produced. Our code works for an arbitrary electron index $p$, which is required because we expect the solution to have $p<2$. Such a detailed treatment of the SSC model is necessary because non-power-law distributions can arise when the electrons are inverse Compton cooled but deep in the Klein-Nishina regime or when $\nu_c<\nu_a$. These cases are precisely the place where we've shown that any solution to the synchrotron model would likely exist. The new methodology is described in detail in appendix \ref{sec:SSC}.

The main goal of the methodology is simple: we give the code a set of free parameters, and it determines the electron distribution required to match the X-ray spectrum of Sw J1644+57 during the first 10 days after its discovery, $f_\nu=0.1 \ \mathrm{mJy}(\nu/1\ \mathrm{keV})^{-0.8}$ for 1 keV$\leq \nu \leq$150 keV. We then calculate the synchrotron and inverse Compton emission expected at different frequencies from the same electron distribution that is producing the X-ray photons. The code will then determine if the following conditions are met: the required luminosity in electrons, protons, and the magnetic field are all less than $10^{50}$ erg/s, and the predicted flux at at non-XRT frequencies are not much larger than the observations at these frequencies. We do not use the optical measurements because the extinction from the host galaxy is uncertain. The uncertain extinction makes it hard to predict an observed optical flux, so instead we use the near-IR observations where the extinction is greatly reduced. We calculate the predicted flux at 0.57 eV to compare to the observed K-band measurements, the predicted integrated flux in the LAT band to compare to the Fermi-LAT upper limits, and the predicted flux at 500 GeV to compare to VERITAS upper limits, accounting for $\gamma+\gamma$ pair opacity for high energy $\gamma$-rays. For frequencies at which a measurement of the flux was obtained, we require the predicted flux from the X-ray producing electrons not to exceed the observed measurement by more than a factor of 2. For frequencies where there are only upper limits on the flux, we require that the predicted flux not exceed these upper limits. We therefore are only trying to account for the X-ray emission while not violating any other observation of Sw J1644+57 or requiring a luminosity greater than $10^{50}$ erg/s, as then the energy requirements would exceed the restmass energy of the tidally disrupted star. For instance, if the predicted K-band flux is much less than the observed flux, that is acceptable because it simply means that the infrared emission from Sw J1644+57 comes from a different source than the X-rays, such as the accretion disk. If the X-ray producing electrons do not violate any observational constraints and require less than $10^{50}$ erg/s of energy, then synchrotron radiation is a plausible mechanism for XRT-BAT emission of Sw J1644+57. 

We show our results in figure \ref{fig:Synch}. We find that a small part of the parameter space survives at $R\lae 2\times10^{13}\ \mathrm{cm}$. Therefore, synchrotron radiation from a shock may be able to produce the early X-ray spectrum observed by Swift. However, we find that the allowed parameter space for such a solution is at small radii, well below the photosphere. Because the electrons are cooled in this region, the injected electron spectrum must be hard, with $p\approx 1.6$. Such a hard spectrum is unexpected in electrons accelerated in shocks. We therefore find that such a model is an unlikely producer of the X-rays in Sw J1644+57.
\section{Proton Synchrotron}\label{sec:Pro_synch_TDE}
In this section we consider proton synchrotron radiation. We find that proton synchrotron model is capable of reproducing the observed X-ray emission of Sw J1644+57 only if the jet is very Poynting dominated  ($\sigma>3\times10^{4}$), and the protons producing keV photons are in the fast cooling regime. The proton synchrotron model is not very efficient, with a radiative efficiency on the order of $1$\%. If the protons are not cooled through synchrotron radiation, then either the proton synchrotron process requires too much energy, or the protons will produce a tremendous amount of $\gamma$-rays through photo-pions. As we show in this section, for proton synchrotron to work the jet must be Poynting dominated, and the protons must be accelerated by magnetic reconnection.

As we have seen in the previous section, preventing a K-band excess is difficult when modeling the observed X-rays of Sw J1644+57. If the X-rays are produced via synchrotron radiation, there are two ways to avoid an excess: either the synchrotron self-absorption frequency is above the optical band, or the cooling frequency, $\nu_c$, is greater than $\sim0.1$ keV. Protons are less radiatively efficient than electrons due to their larger mass. The radiative inefficiency of protons means that synchrotron self-absorption cannot prevent a K-band excess in proton synchrotron models. Following a similar argument as the one used in the synchrotron model section, the maximum synchrotron self-absorption frequency from protons radiating in a magnetic field $B'$ is smaller than electrons self-absorption frequency by a factor $\left(m_p/m_e\right)^{2/(5+2\beta)}.$ The ratio between the two self-absorption frequencies is a factor of $\sim10$ if $\beta=0.8$. We showed the difficulty in making $\nu_a$ large enough for electron synchrotron models \S \ref{sec:Synch}, and the fact that $\nu_a$ is a factor of 10 smaller for proton synchrotron means that $\nu_a$ will not be large enough unless $R\sim 10^{12}\ \mathrm{cm}$. Therefore, the proton synchrotron model only works if $\nu_i\sim 1$ keV and $\nu_c\gae 0.1$ keV. 

Protons with Lorentz factor $\gamma_i$ in a magnetic field $B'$, will radiate at a frequency
\begin{equation}
\nu_i=6.3\times10^{-15}B'\Gamma\gamma_i^2(1+z)^{-1}\ \mathrm{keV}.
\end{equation}
If there are an isotropic equivalent number of protons, $N_p$, radiating at this frequency, the observed flux will be
\begin{equation}
f_{\nu_i}=0.1\ \mathrm{mJy}\ N_{p,55}B'\Gamma(1+z)/d_{L,28}^2
\end{equation}
Rewriting the above equations in terms of the luminosity carried by the magnetic field, $\mathcal{L}_B$, eq (\ref{eq:MagLum}), we find 
\begin{equation}\label{eq:pro_gamma_i}
\gamma_i\sim8\times10^4\ \nu_{i,\mathrm{keV}}^{1/2} \mathcal{L}_{B,47}^{-1/4}R_{14}^{1/2}(1+z)^{1/2},
\end{equation}
and
\begin{equation}
N_p=4\times10^{51}f_{\nu,\mathrm{mJy}}\mathcal{L}_{B,47}^{-1/2}R_{14}d_{L,28}^2/(1+z)
\end{equation}

The requirement that the proton cooling frequency be equal to or greater than 1 keV can be easily satisfied because the synchrotron cooling time for protons is larger than the cooling time for electrons by a factor $(m_p/m_e)^3$.
\begin{eqnarray}
t'_{c,p}(\gamma_i) &=&\frac{6\pi m_p^3 c}{m_e^2\sigma_TB'^2\gamma_i}\\ 
&=& 9\times10^4\ \nu^{-1/2}_{i,\mathrm{keV}}\mathcal{L}_{B,47}^{-3/4}R_{14}^{3/2}\Gamma^2(1+z)^{-1/2}\ \mathrm{s}.
\end{eqnarray}
The proton synchrotron cooling frequency is
\begin{equation}\label{eq:nu_c_pro}
\nu_c=3\times10^3\ \mathcal{L}_{B,47}^{-3/2}R_{14}(1+z)^{-1} \ \mathrm{keV}.
\end{equation}

We consider two regimes, one where the protons are slow cooling, $\nu_c>150$ keV, and one where the protons are fast cooling, $\nu_c\lae1$ keV. First we consider the slow cooling regime.

In the slow cooling regime the energy carried by the protons is quite large. The proton luminosity is
\begin{align}
\mathcal{L}_p&=\frac{2\Gamma^3\gamma_i m_p c^3 N_p}{R}\\
\mathcal{L}_p &\approx3\times10^{50}\ \Gamma^3\mathcal{L}_{B,47}^{-3/4}f_{\nu,\mathrm{mJy}}\nu^{1/2}_{i,\mathrm{keV}} \frac{R_{14}^{1/2}d_{L,28}^2}{(1+z)^{1/2}}\ \mathrm{erg/s}
\end{align}
For fiducial values $\mathcal{L}_p\approx 9\times10^{48} \ \Gamma^3\mathcal{L}_{B,47}^{-3/4}R_{14}^{1/2}$ erg/s. If we assume the observed variability is due to the dynamical time of the jet,  $\mathcal{L}_p\approx 9\times10^{50} \ \mathcal{L}_{B,47}^{-3/4}R_{14}^{2}\delta t_2^{-3/2}$ erg/s. This luminosity is too large unless $\mathcal{L}_{B}$ is significantly larger than the observed X-ray luminosity or $R<10^{14}\ \mathrm{cm}$. If $\mathcal{L}_B$ is much larger than the observed X-ray luminosity, there will be a cooling break in the BAT band. So the only option is to decrease $R$. As we now show, in proton synchrotron models $R$ must be greater than $10^{14}$ cm, because when $R<10^{14}$ cm, photo-pion emission will dominate. We calculate the photo-pion emission in a similar manner as \citet{2013MNRAS.429.3238C}. 

In the photo-pion process a photon interacts with a proton to cause a resonance that decays into pions. The resonance with the largest cross section is the $\Delta^+$ resonance, $p^+ + \gamma\rightarrow\Delta^+$. The $\Delta^+$ resonance quickly decays via the strong force into pions that further decay into photons, muons, electrons, and neutrinos.  
A proton with a Lorentz factor $\gamma_i$  will undergo a photo-pion collision with a photon with an observed frequency $\nu_{p\gamma}$ if
\begin{equation}
\gamma_i \frac{h\nu_{p\gamma}(1+z)}{\Gamma}\geq 200\ \mathrm{MeV}
\end{equation}
Using eq (\ref{eq:pro_gamma_i}) for $\gamma_i$ we find that the characteristic photon energy for the photo-pion process is
\begin{equation}
\nu_{p\gamma}=2.5\ \mathrm{keV}\ \Gamma\nu_{i,\mathrm{keV}}^{-1/2}\mathcal{L}_{B,47}^{1/4}R_{14}^{-1/2}(1+z)^{-3/2},
\end{equation}
If $\nu_{p\gamma}\lae 10$ keV, we can approximate the number density of photons with frequencies $\gae \nu'_{p\gamma}$  in the comoving frame is 
\begin{equation}
n'_{p\gamma} \sim \frac{U'_X}{h\nu'_{p\gamma}} = \frac{L_X}{4\pi R^2 \Gamma^2 c h\nu'_{p\gamma}}
\end{equation}
\begin{equation}
n'_{p\gamma} =7\times10^{15} L_{X,47}\Gamma^{-2}R_{14}^{-3/2}\nu_{i,\mathrm{keV}}^{1/2}\mathcal{L}_{B,47}^{-1/4}(1+z)^{1/2}
\ \mathrm{cm^{-3}}
\end{equation}
The optical depth to photo-pion process is $\tau_{p\gamma} = n_{p\gamma}\sigma_{p\gamma}R/\Gamma$, where $\sigma_{p\gamma}$ is the cross section of delta resonance, $\sigma_{p\gamma}=5\times10^{-28}\ \mathrm{cm^2}$. 
\begin{equation}
\tau_{p\gamma}\sim350 L_{X,47}R_{14}^{-5/2}\Gamma^{-3}\nu_{i,\mathrm{keV}}^{1/2}\mathcal{L}_{B,47}^{-1/4}(1+z)^{1/2}
\end{equation}
Constraining $\Gamma$ by the variability time we find
\begin{equation}
\tau_{p\gamma} =5\ L_{X,47}R^{-4}_{14}\nu_{i,\mathrm{keV}}\mathcal{L}^{-1/4}_{B,47}(1+z)^{-1}\delta t_2^{-3/2}
\end{equation}
A proton will lose $\sim$20\% of its energy in a single photo-pion interaction. Since the optical depth to photo-pion scatterings is larger than 5 when $R\lae 10^{14}$ cm, nearly all of the luminosity carried by the protons will be put into photo-pions. The photons produced in the photo-pion interaction will be very high energy and not able to escape the jet due to $\gamma+\gamma$ pair opacity. However, the energy will cascade down until it is able to escape the jet releasing $\gae10^{49}$ erg/s worth of luminosity in the $\gamma$ rays. A photo-pion bump is not seen, so the slow cooling regime can be ruled out.

We now consider the fast-cooling regime. From eq (\ref{eq:nu_c_pro}), it is clear that $\nu_c$ will less than 1 keV if $\mathcal{L}_{B} \gae 10^{49}$ erg/s and $R\approx 10^{14}$ cm. When the protons are cooled by the synchrotron process, all their energy will come out in X-rays, so
\begin{align}
\mathcal{L}_p&\approx\Gamma^2\gamma_i m_p c^2N_p/t'_{c,p}(\gamma_i)\\ 
 &\approx 5\times10^{48}f_{\nu_i,\mathrm{mJy}}\nu_{i,\mathrm{keV}}d_{L,28}^2\ \mathrm{erg/s}\\
 &\sim 3\times10^{47}\ \mathrm{erg/s}.
\end{align}
For the protons to be cooled, the jet must be strongly Poynting dominated. To calculate the magnetization parameter, $\sigma$, we assume that there are $\eta_p$ cold protons for every proton that is accelerated to $\gamma_i$, 
\begin{align}
\sigma &= \frac{\mathcal{L}_B}{\mathcal{L}_{p,\mathrm{cold}}}=\frac{\gamma_i}{\eta_p}\frac{\mathcal{L}_B}{\mathcal{L}_p} \\
\sigma &= 5\times10^4\ \mathcal{L}_{B,49}^{3/4}\eta_p^{-1}R_{14}f_{\nu_i,\mathrm{mJy}}^{-1}d_{L,28}^{-2}(1+z)^{3/2}
\label{eq:sigma_pro}
\end{align}
$\sigma \approx 2\times10^{6}\eta_p^{-1}\mathcal{L}_{B,49}^{3/4}R_{14}$ for the fiducial values. Such a large value of sigma will prevent any strong shocks from forming, so shocks will not be able to accelerate the protons. However, magnetic reconnection is be able to accelerate the protons, as long as the energy given to the protons not exceed the initial energy in the magnetic field, or $\gamma_i\leq\sigma$. Using eq (\ref{eq:pro_gamma_i}), when $\mathcal{L}_{B}=10^{49}$ erg/s, $\gamma_i\approx 3\times10^{4}R_{14}^{1/2}$ and $\sigma\approx 2\times10^{6}\eta_p^{-1}R_{14}$. $\gamma_i$ is less than $\sigma$ as long as $\eta_p$ is $\leq100$.

Therefore, the proton synchrotron process can match the X-ray spectrum of Sw J1644+57 if the jet is Poynting dominated with an efficiency $\sim 1$\%. The minimum Lorentz factors of the protons is $\gae 2\times10^{4}$. The magnetization of the jet, $\sigma$ must be larger than $\gamma_i$, or $\sigma\gae 2\times10^{4}$. In the proton synchrotron model, the jet is very Poynting dominated.
\section{Internal Inverse Compton Radiation}
\label{sec:IC}

In this section we consider the possibility that the X-ray emission observed during Sw J1644+57 is produced by Compton scattering lower energy photons that are internal to the jet itself. External inverse Compton is considered in the next section. We show that inverse Compton radiation is not capable of producing the observed averaged keV flux of Sw 1644+57 without causing an excess in a different wavelength or requiring too much energy. 

In inverse Compton radiation, a seed photon of frequency \(\nu_s\) and flux $f_{\nu_s}$ are scattered into the keV band by electrons with LF $\gamma_\mathrm{IC}$ and with Compton Y of $\widetilde{Y}_\mathrm{IC}=\gamma_\mathrm{IC}^2\tau_\mathrm{IC}$, where \(\tau_\mathrm{IC}=N_e\sigma_T/(4\pi R^2)\), and $N_e$ is the number of electrons with LF greater than $\gamma_\mathrm{IC}$ that participate in the IC radiation. We can relate the seed photon field to the observed field by using
\begin{equation}\label{eq:IC_stuff}
\nu_\mathrm{IC}\approx \gamma_\mathrm{IC}^2\nu_s\approx 1\ \mathrm{keV}; \quad\quad 
f_{\nu, \mathrm{IC}}\approx \tau_\mathrm{IC} f_{\nu_s}\approx 0.1\ \mathrm{mJy}
\end{equation}

In this section, we can constrain the parameter space allowed for inverse Compton radiation producing the observed X-rays. We try to make as few assumptions as possible about the seed photons. In the inverse Compton process, there are 6 free parameters: $f_{\nu_s}$, $\nu_s$, $\gamma_\mathrm{IC}$, $Y$, $R$, $\Gamma$. $f_{\nu_s}$ and $\nu_s$ are the seed photon flux and frequency in the observer frame. These are left completely free and are solved for by matching the observed flux of $\sim$ 0.1 mJy at 1 keV. $\gamma_\mathrm{IC}$ is the Lorentz factor of electrons that scatter the seed field to 1 keV. $\widetilde{Y}=\gamma_\mathrm{IC}^2\tau_\mathrm{IC}$ is a quantity closely related to Compton $Y$, and as before, $R$ is the radius of emission and $\Gamma$ is the bulk Lorentz factor.

We assume that the seed flux is produced through synchrotron radiation from electrons that may have a different Lorentz factor than $\gamma_\mathrm{IC}$. Additionally, we do not require that the electrons producing the seed field be part of the same power-law distribution as the electrons with Lorentz factors $\sim \gamma_\mathrm{IC}$. For the X-ray spectral index to match the observed value of $-0.8$, the seed flux spectrum above $\nu_s$ must be as soft or softer than observed spectrum of $\nu^{-0.8}$. Therefore, the self-absorption frequency must be less than or equal to $\nu_s$. When $\nu_s$ is below the K-band, synchrotron self-absorption from the electrons producing the seed field cannot change the K-band flux predicted from the inverse Compton scattered photons. $\nu_s$ is below the K-band when $\gamma_\mathrm{IC}\gae25$. As we will now show, if $\gamma_\mathrm{IC}\gae 25$, inverse Compton process cannot be responsible for the keV flux observed in Sw J1644+57.

\begin{figure*}

\includegraphics[width=\textwidth]{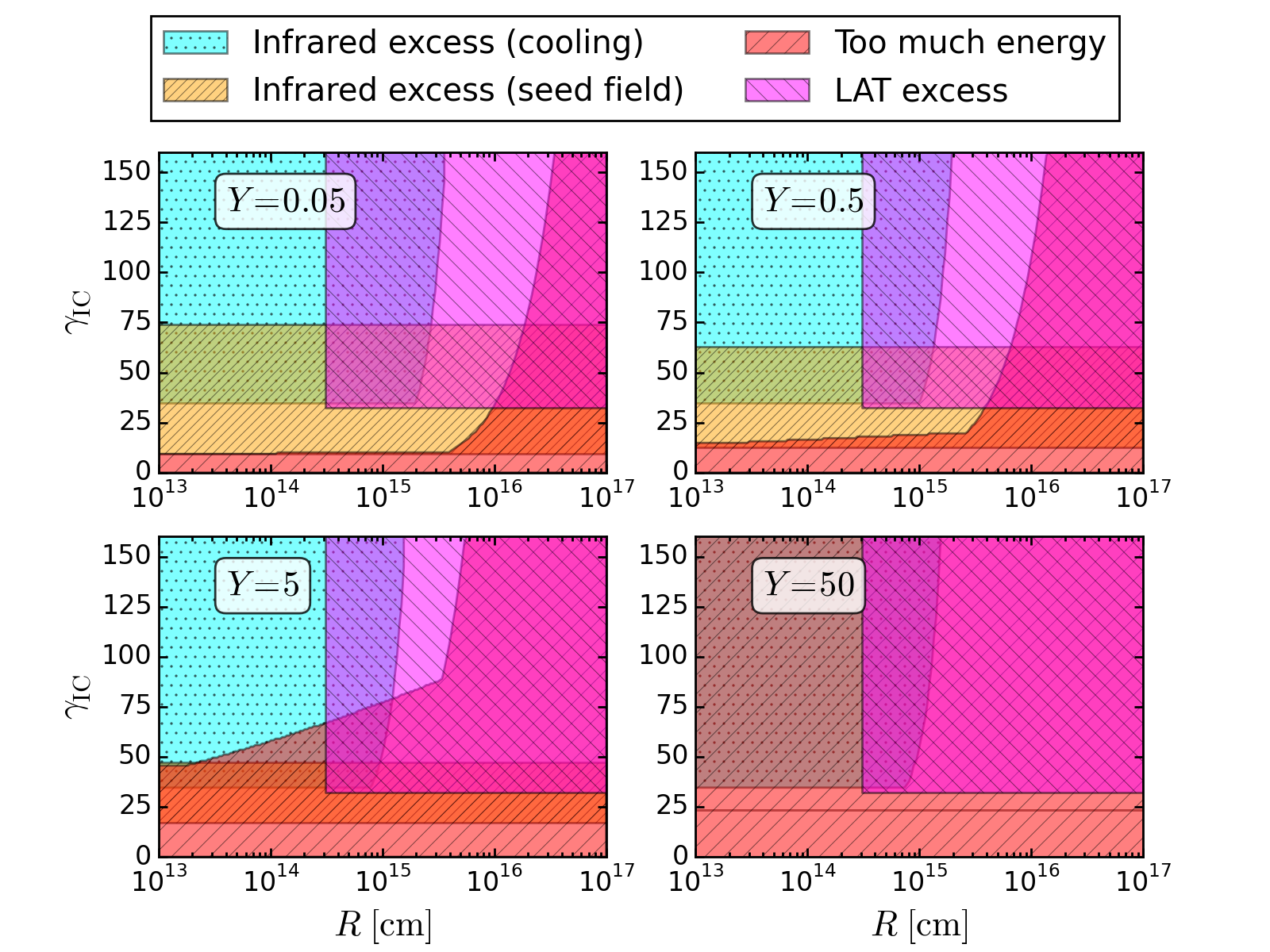}
\caption[The available parameter space to the inverse Compton model of Sw J1644+57]{This figure shows the various constraints the observations of Sw J1644+57 impose on any internal inverse Compton model (see \S \ref{sec:IC} for details). The $y$-axis is $\gamma_\mathrm{IC}$, the electrons that scatter a seed photon field to 1 keV. The $x$-axis is $R$, the radius of emission. The different panels show different values of Compton $Y$. Infrared excess (cooling) is the area where cooled electrons will IC scatter the seed field into the infrared and overproduced the observed K-Band flux. Infrared excess (seed field) is the area where the required flux in the seed field that is up-scattered to 1 keV exceeds the observed K-band measurements. LAT excess is where photons of $\sim 150$ keV would be up-scattered into the LAT band and violate the strict upper limits measured in Sw J1644+57. Too much energy where the energy required is greater than the rest mass of a solar mass star (a luminosity $\gae 10^{50} \mathrm{erg/s}$). Overlapping regions are where the internal IC model will not work for several reasons. Since there is no white space in this figure, the internal IC model is ruled out.}
\label{fig:IC}
\end{figure*}

The constraints outlined in this section are shown graphically in figure \ref{fig:IC}. First we show that when $R$ is less than a few times $10^{15}$ cm and  $\gamma_\mathrm{IC}\geq25$, the electrons will cool rapidly by inverse Compton scattering X-ray photons. The cooled electrons will then overproduce in the K-band because they will scatter the seed field photons to energies $\sim 1 \mathrm{eV}$.  

The inverse Compton power from an electron with Lorentz factor $\gamma$ traveling through a photon field with energy density $U'_{\gamma}$ can be approximated by breaking the photon field into 2 parts: $U'_s$, the seed field that gets boosted from frequency $\nu_s$ to $\nu_\mathrm{IC}\sim$ 1 keV, and the observed X-ray flux integrated over the XRT and BAT frequencies, $f_{\nu_x}$ using eq (\ref{eq:pow_IC}):
\begin{align}
P_\mathrm{IC} &\approx P_{\mathrm{IC},s} + P_{\mathrm{IC},X} \\
P_\mathrm{IC} &=
\frac{4}
{3}\sigma_T c\gamma^2U'_s+
\frac{12\sigma_T d_L^2}{\left(\Gamma\gamma^2R\right)^2}
\int
{d\nu_x f_{\nu_x} F_\mathrm{IC}(\epsilon_{\nu_x}, \gamma)},
\end{align}
Ignoring Klein-Nishina effects and using $U'_s\approx U'_X/\widetilde{Y}$ where $U'_X$ is the energy density in X-ray photons in the co-moving rest frame, the inverse Compton power is approximately,
\begin{equation}
P_\mathrm{IC}\approx \frac{4}{3}\sigma_T c \gamma^2U_X\left(1+\frac{1}{\widetilde{Y}}\right)
\end{equation}
Re-writing the above equation in terms of the observed X-ray luminosity $U'_X=L_x/(4\pi cR^2\Gamma^2)$ the cooling time becomes
\begin{equation}\label{eq:t_cool_IC}
t'_{c,\mathrm{IC}}\sim 120\ s\frac{R_{14}^2\Gamma^2_1}{\gamma L_{X,47}}\left(1+\frac{1}{ \widetilde{Y}}\right)^{-1} 
\end{equation}
Comparing the inverse Compton cooling time to the dynamical time, \( t'_\mathrm{dyn}=170 R_{14}\Gamma_1^{-1}\), it is clear that even electrons with Lorentz factor order 2 will cool quickly via inverse Compton radiation at $R=10^{14}\ \mathrm{cm}$. The cooling Lorentz factor is $\gamma_{c, \mathrm{IC}}$, and its corresponding inverse Compton frequency is
\begin{equation}
\gamma_{c,\mathrm{IC}}\approx 7
\frac{R_{15}\Gamma_1^3}{L_{X,47}}
\left(1+\frac{1}{\widetilde{Y}}\right)^{-1} ; \quad
\nu_{c, \mathrm{IC}}=
\nu_\mathrm{IC}\left(\frac{\gamma_{c,\mathrm{IC}}^2}{\gamma_\mathrm{IC}^2}\right)
\end{equation}
The inverse Compton flux between $\nu_{c,\mathrm{IC}}$ and $\nu_\mathrm{IC}$ is proportional to $\nu^{-1/2}$, and the inverse Compton flux below $\nu_{c,\mathrm{IC}}$ is proportional to $\nu^1$ \citep{Blumenthal1970} . When $\gamma_{c,\mathrm{IC}}<\gamma_\mathrm{IC}$, the extrapolated K-band flux will be
\begin{align}
f_\mathrm{K-band}&\approx f_{\nu,\mathrm{IC}}\left(\frac{\nu_{c,\mathrm{IC}}}{\nu_\mathrm{IC}}\right)^{-1/2}\left(\frac{\nu_\mathrm{K-band}}{\nu_{c,\mathrm{IC}}}\right)\\
& \approx  5.7\times10^{-4} f_{\nu,\mathrm{IC}}\frac{\gamma_\mathrm{IC}^3}{\gamma_{c, \mathrm{IC}}^3}
\end{align}
The measured flux in the K-band at early times was 0.15 mJy, and the measured flux at 1 keV was $\sim 0.1$ mJy. We find that $\gamma_\mathrm{IC}\leq 17 \gamma_{c,\mathrm{IC}}$; otherwise the K-band flux will be too large. Therefore
\begin{equation}\label{eq:IC_gamma_max}
\gamma_\mathrm{IC}\leq \max{\left[25,\ 12\ \frac{R_{14}\Gamma_1^3}{L_{X,47}}\left(1+\frac{1}{\widetilde{Y}}\right)^{-1}\right]}
\end{equation}
The region disallowed by eq (\ref{eq:IC_gamma_max}) corresponds to the dotted cyan area in figure \ref{fig:IC}.

However, when \(12\ R_{14} \Gamma_1^3 L_{X,47}^{-1} \gae 25\), the inverse Compton process will not work due to overproduction in the LAT band. This constraint corresponds to backslashed magenta area in figure \ref{fig:IC}. The optical depth at 100 MeV can be approximated using the total X ray luminosity and eq (\ref{eq:tau_gammagamma}) and (\ref{eq:gammagamma_min}). The photons with an observed energy of 100 MeV will pair produce with photon with energy
\begin{equation}
\nu_{\gamma\gamma}\sim\ 2\Gamma^2(1+z)^{-2}\nu_{obs,100\ \mathrm{MeV}}^{-1} \ \mathrm{keV}
\end{equation}
With $\Gamma\sim 10$ pair production will occur with photons with energies in the BAT band. Since the observed spectral index is $\sim1$ to 150 keV, we can estimate the number of photons at $\nu_{\gamma\gamma}$ as
\begin{align}
n'_{\gamma\gamma} & \sim  \frac{L_X}{4\pi R^2 c \Gamma^2 h\nu'_{\gamma\gamma}}\\
n'_{\gamma\gamma} & \sim  8\times10^{15}\ L_{X,47}R_{14}^{-2}\Gamma^{-3}(1+z)
\ \mathrm{cm^{-3}}
\end{align}
Which has a corresponding optical depth of
\begin{equation}
\tau_{\gamma\gamma}= n'_{\gamma\gamma}\sigma_TR/\Gamma\sim 2L_{X,47}R_{14}^{-1}\Gamma_1^{-4}
\end{equation}
If the electrons are uncooled, \(12 R_{14}\Gamma_1^3L_{X,47}^{-1}\gae 25\) implies that the optical depth to pair production for photons of energies $\sim 100$ MeV must be less than
\begin{equation}
\tau_{\gamma\gamma}\lae \Gamma_1^{-1}
\end{equation}
The optical depth of pair production to 100 MeV photons is less than one, because $\Gamma_1$ is greater than 1 when $R_{14}\Gamma_1^3\gae20$ and $t_\mathrm{dyn}<100$ seconds. Therefore photons with energies $\gae 100$ MeV will escape the jet. The emission would over produce the stringent upper limits from the LAT unless $\widetilde{Y}\lae4\times10^{-2}$ (\textit{i.e.}, the ratio of the X-ray luminosity in the XRT band to the upper limits from \textit{Fermi}-LAT). However such a low Compton $Y$ can be disregarded due to the large luminosity required in the seed field of photons. Therefore, as can been seen in the overlapping magenta and cyan areas in figure \ref{fig:IC}, it is a strict requirement that
\(\gamma_\mathrm{IC} \lae  25\).

When $\gamma_\mathrm{IC}\lae25$, the seed frequency, $\nu_s$, is greater than K-band frequency. It is therefore possible that the seed photons that get upscattered into the XRT band will overproduce in the K-band. The area where this happens is shown in figure \ref{fig:IC} as the goldenrod colored region with dense forward slashes.

In synchrotron spectra, the hardest spectral index achievable is approximately $f_\nu\propto \nu^{5/2}$. Using eq (\ref{eq:IC_stuff}), the minimum flux at 0.57 eV can be calculated using the observed flux at 1 keV, \(f_{\nu,\mathrm{IC}}\)
\begin{align}
f_\mathrm{K-band}
&\approx \frac{f_{\nu,\mathrm{IC}}}{\tau_\mathrm{IC}}\left(\frac{0.57\ \mathrm{eV}}{\nu_\mathrm{IC}/\gamma_\mathrm{IC}^2}\right)^{5/2}\\
f_\mathrm{K-band} & \sim  8\times10^{-8}\gamma_\mathrm{IC}^7\widetilde{Y}^{-1}f_{\nu,\mathrm{IC}}\nu_\mathrm{IC,keV}^{-5/2}
\end{align}
Requiring $f_\mathrm{K-band}\lae 2f_{\nu,\mathrm{IC}}$ to avoid an infrared excess, we come to the tight constraint on $\gamma_\mathrm{IC}$,
\begin{equation}\label{eq:infrared_excess}
\gamma_\mathrm{IC}\lae 11\ \widetilde{Y}^{1/7}
\end{equation}

The final constraint is that the energy required for the process to work is not too large. The total number of electrons required in the jet is
\begin{equation}
N_{e,\mathrm{tot}}=\frac{4\pi R^2\widetilde{Y}}{\sigma_T\gamma_\mathrm{IC}^2}\max{\left(1,\frac{t'_\mathrm{dyn}}{t'_{c,\mathrm{IC}}}\right)}
\end{equation}
As we showed earlier, it is likely to be the case that the electrons are cooled. Using eq (\ref{eq:t_cool_IC}) for the inverse Compton cooling time we find that the total number of electrons in the jet must be greater than
\begin{equation}\label{eq:IC_Ne_tot}
N_{e,\mathrm{tot}}\gae 1.5\times10^{53} \frac{L_{X,47}t'_\mathrm{dyn}}{\Gamma^2\gamma_\mathrm{ IC}}\left(\widetilde{Y}+1\right).
\end{equation}
Assuming $\eta_p$ cold protons per hot electron in the jet, the luminosity carried by the protons in the jet is 
\begin{equation}
\mathcal{L}_p\approx\eta_p\Gamma^2 m_p c^2 N_{e,\mathrm{tot}}/t'_\mathrm{dyn}.
\end{equation}
Using eq (\ref{eq:IC_Ne_tot}) for $N_{e,\mathrm{tot}}$ relation of
\begin{equation}
\mathcal{L}_p\gae 2.3\times10^{50} \ \eta_p\frac{L_{X,47}}{\gamma_\mathrm{IC}}\left(\widetilde{Y}+1\right)
\ \mathrm{erg/s}.
\end{equation}
Using the inequality for $\gamma_\mathrm{IC} $ in eq (\ref{eq:infrared_excess})
\begin{equation}
\mathcal{L}_p\gae 2\times10^{49}\ \eta_p L_{X,47}\left(\widetilde{Y}^{6/7}+\widetilde{Y}^{-1/7}\right)
\end{equation}
The mean luminosity of Sw J1644+67 for the first 10 days is $\approx 3\times10^{47}$ erg/s. Using this luminosity we find $\mathcal{L}_p\gae 9\times10^{49}$ erg/s for the first $10^6$ seconds of Sw 1644+57. This energy requirement is prohibitive even if $\eta_p\sim1$. The region where the energy required is too large is shown in figure \ref{fig:IC} as the red forward-slashed region. The ruled out regions overlap in figure \ref{fig:IC} with no space that is not ruled out. Therefore, the internal inverse Compton radiative mechanism can be ruled out for Sw J1644+57. 

\def\spa{{\mbox{ }}}

\section{External Inverse Compton Radiation}
\label{sec:EIC}
In this section, we consider the possibility that the X-rays may come from an external inverse Compton (EIC) process, i.e., the up-scattering of external radiation field photons by electrons traveling inside of the jet. The EIC process will boost soft photons' energy by a factor of $\Gamma^2\gamma_e^2$. We only consider the up-scattering of photons that are produced in an optically thick wind that comes off of the disk. This is in contrast to \citet{Bloom2011}, where the authors fit the X-ray data with an EIC model in which electrons in the jet up-scatter photons produced in the disk. However, the disk size  is restricted to $\sim R_T = 7\times 10^{12} m_6^{1/3}\ \rm cm$ for a typical star like the Sun \citep{2009MNRAS.400.2070S}.\footnote{Disks expand as angular momentum is transported outwards, but in the TDE context, the total disk angular momentum is small and the disk expands to the circularization radius $2R_p \sim R_T \sim 10^{13}\ \mathrm{cm}$.}
The disk expands to a radius larger than the self-shielding radius (see eq \ref{eq:8}), and in the absence of an optically thick wind, the jet may be dragged to a halt due to the IC scattering of radiation from the disk. This is because at $R=10^{13}$ cm, disk photons will penetrate the jet in the transverse direction and the IC power of each electron in the jet is $P_{\rm IC} = L_{\rm disk} \sigma_T \Gamma^2 \gamma_e^2 /(4\pi R^2$). Compare this to the amount of kinetic energy shared by each electron is $\Gamma m_p c^2$ (Note: $m_p$ here instead of $m_e$ because most of the jet's momentum is carried by protons). Therefore, the EIC drag timescale is $\Gamma m_p c^2/P_{\rm IC}$. Compare this to the dynamical timescale ($t_{\rm dyn} = R/(2c)$):
\begin{equation}
\label{eq:EIC_cool_ratio}
\frac{t_{\rm EIC}}{t_{\rm dyn}} =
1.7 \frac{R_{13}}{L_{\rm disk,45} \Gamma_1 \gamma_e^2}
\end{equation}
from the above equation we can see that the EIC drag is strong. An optically thick wind, which naturally comes out of a super-Eddington disk, alleviates this IC drag problem. In a TDE, the jet is expected to be surrounded by an optically thick wind loaded with radiation from the disk \citep{2011ApJ...736....2O, 2014ApJ...796..106J} and photons from the disk are scattered (multiple times) by electrons in the wind before they arrive at the relativistic jet. This process is described in detail below.

Photons in the optically thick wind launched by the super-Eddington disk are advected by electron scattering and cool adiabatically to $\sim 10$ eV before escaping. For Sw J1644+57, if we assume that the jet has bulk Lorentz factor $\Gamma=10$, and that the electrons are cold, $\gamma_e=1$, the scattered photons will have energy $\sim \Gamma^2 \gamma_e^2$ larger, $\sim  1\  {\rm keV}$. Therefore, EIC emission is a plausible way of producing the observed X-rays in Swift J1644+57. The EIC model for another relativistic TDE Sw J2058+05 is described by \citet{2016arXiv160201468L}. For completeness, we reproduce their calculations and test if the model is consistent with observations in Sw J1644+57. First we describe the characteristics of the jet and the radiation from the wind. Then we calculate the expected EIC luminosity and spectrum in the XRT-BAT band. We find that the EIC model is marginally consistent with the observations if the efficiency in launching the wind is large, and the Lorentz factor of the jet is not too large, $\Gamma\lae5$. 

\subsection{Jet Characteristics}
We assume a baryonic (pure-Hydrogen) jet with narrow half opening angle, $\theta_j\ll1$. Electron number density is denoted  as $n_e$ (SMBH rest frame) and $n_e'$ (jet comoving frame).  The optical depth of the jet for an external photon traveling in the transverse direction is
\begin{equation} 
  \label{eq:10}
  \tau_{\rm trvs} = \frac{\pi \theta_j^2 R^2 \Delta R n_e \sigma_T}{2\pi
  R \theta_j \Delta R}  = \frac{1}{2}R\theta_j n_e \sigma_T
\simeq 0.59 \frac{L_{j,48}\theta_{j,-1}}{R_{13}\Gamma_1},
\end{equation}
and in the radial direction towards the SMBH is
\begin{equation}
  \label{eq:9}
      \tau_{r}  = R n_e \sigma_T = 11.7\frac{L_{j,48}}{R_{13}\Gamma_1}
\end{equation}
We can see that it's easier for external photons to penetrate the jet in the transverse direction than the radial direction. Defining the ``self-shielding radius'' where $\tau_{trvs}=1$, we have 
\begin{equation}
  \label{eq:8}
  R_{j,{\rm self}} = 5.9 \times 10^{12} 
  \frac{L_{j,48}\theta_{j,-1}}{\Gamma_1} \ {\rm cm}
\end{equation}

The optical depth for a photon traveling along with the jet is $\tau_{\rm jet}$, defined in eq (\ref{eq:tau_jet}). Note that $\tau_{\rm jet}$ is much smaller than $\tau_{\rm trvs}$ (by a factor of $2/\Gamma^2\theta_j$), so we can ignore multiple EIC scatterings at radii $R>R_{j,{\rm self}}$.

\subsection{Radiation from the Wind}\label{wind_radiation}
The gas from the tidally disrupted star, falls back initially in elliptical orbits, shocks on itself, and circularizes near the pericenter of the star's orbit. The gas then settles in a disk around the black hole. The shocks as well as the disk's viscosity convert orbital kinetic energy to thermal energy. When the fallback is super-Eddington, the gas density is large and thermal energy is quickly converted to radiation energy, mostly via free-free emission. However, the radiation is trapped by electron scattering, and therefore the shocked gas is radiation pressure dominated. While the gas accretes inward and releases gravitational energy, it is likely that some fraction of the mass is unbound and leaves the accretion disk in a wind. Although The wind launching process still has large theoretical uncertainties \citep{2011ApJ...736....2O, 2012MNRAS.426.3241N, 2014ApJ...796..106J}, we assume the wind is radiation driven and construct a simple 1-D model of the radiation characteristics from the wind in this section.

Generally, we use upper case $R$ to denote the true radius (in cm) and the lower case $r$ for the dimensionless radius $R/R_S$ (normalized by the Schwarzschild radius $R_S = 3\times 10^{11} M_{\bullet,6}\ \mathrm{cm}$). Also, the true accretion, outflowing (sub ``w''), and fallback (sub ``fb'') rates (in $M_\odot \spa {\rm yr^{-1}}$) are denoted as upper case $\dot{M}$ and the dimensionless rates are normalized by the Eddington accretion rate as $\dot{m} = \dot{M}/\dot{M}_{\rm Edd}$. The Eddington accretion rate is defined as $\dot{M}_{\rm Edd} = 10L_{\rm Edd}/c^2$,  and $L_{\rm Edd} =1.5\times10^{44} M_{\bullet, 6}\ \mathrm{erg\ s^{-1}}$, where $M_{\bullet, 6}$ is BH mass in $10^6M_\odot$.

For a star with mass $M_* = m_* M_\odot$ and radius $R_* = r_* R_\odot$, the (dimensionless) tidal disruption radius is
\begin{equation}
  \label{eq:1}
  r_T = \frac{R_*}{R_S} \left( \frac{M_{\bullet}}{M_*} \right)^{1/3}
  \simeq 23.3\ M_{\bullet,6}^{-2/3} m_*^{-1/3} r_*
\end{equation}
For the star to be tidally disrupted, the star's original orbit must have pericenter distance $r_p<r_T$. We assume that the wind is launched at close to the escape velocity at a radius $r_o\simeq 2r_p$ (i.e. circularization radius), where the radiation energy and kinetic energy are in equipartition:
\begin{equation}
  \label{eq:11}
  aT_o^4\simeq \frac{1}{2} \rho_w v_w^2
\end{equation}
where $a$ is the radiation constant, $T_o$ is the temperature at $r_o$, $\rho_w$ is wind density, and $v_w\simeq v_{\rm esc}(r_o)\simeq r_o^{-1/2}c$ is the outflowing velocity. Following \citet{2009MNRAS.400.2070S}, we assume a fraction $f_{\rm out}$ of the fallback gas is leaves in the wind (assuming spherical geometry), so we have 
\begin{equation}
  \label{eq:13}
  4\pi r_o^2 \rho_w(r_o) v_w = \dot{M}_w = f_{\rm out} \dot{M}_{\rm fb} 
\end{equation}
where $\dot{M}_{\rm fb}$ is the mass fallback rate. 

For a thick disk, we expect the viscous time to be much shorter than the fallback time, so the BH accretion rate is $\dot{M}\simeq (1-f_{\rm out}) \dot{M}_{\rm fb}$ (assuming the mass carried away by the jet is negligible). If a fraction $f_{j}$ of the accretion power $\dot{M}c^2$ is used to power the jet, we have
\begin{equation}
  \label{eq:3}
  L_j(1-\cos{\theta_j}) \simeq f_j (1-f_{\rm out}) \dot{M}_{\rm fb} c^2  \simeq
  \frac{f_j (1-f_{\rm out})}{f_{\rm out}} \dot{M}_{w} c^2
\end{equation}
where $L_j$ is the isotropic jet power, and $\theta_j$ is the half opening angle. For example, a set of values \{$f_j = f_{\rm out} = 0.5$, $\theta_j= 0.1$\} yields $\dot{m}_w = 6.7L_{j,48} M_{\bullet,6}^{-1}$. Since the BH's rotation energy may play some role in launching the jet \citep{1977MNRAS.179..433B, 2011MNRAS.418L..79T}, $f_j>1$ may be allowed and we leave the jet power $L_{j}$ as a free parameter.

From eq (\ref{eq:11}) and (\ref{eq:13}), we can get the radiation temperature at the wind launching site
\begin{equation}
  \label{eq:17}
T_o\simeq 1.3\times 10^{6} r_o^{-5/8} M_{\bullet,6}^{-1/4} \dot{m}_w^{1/4}
 \ \mathrm{K}
\end{equation}
The opacity is dominated by electron scattering \citep{2005MNRAS.360..458B, 2015MNRAS.447L..60S}, so the optical depth of the $>R$ region is 
\begin{equation}
  \label{eq:4}
  \tau_w  =  \rho_w \kappa_s R \simeq 4.7 \dot{m}_{w,1} r_{o,1}^{1/2}
  R_{13}^{-1}
\end{equation}
where $\kappa_s = 0.2(1+X) \ \mathrm{cm^2\ g^{-1}}$ is the opacity from Thomson scattering. We choose a Hydrogen mass fraction of $X=0.7$, but the results are not sensitive to our choice in $X$. The wind becomes transparent ($\tau_w=1$) at radius 
\begin{equation}
  \label{eq:5}
  r_{w, {\rm tr}} \simeq 1.6\times10^2 \dot{m}_{w,1} r_{o,1}^{1/2}
\end{equation}
Below $r_{w, {\rm tr}}$, photons escape by diffusion. The radius where diffusion time equals to the dynamical time (i.e., $\tau_w=c/v_w$) is called the ``advection radius'', $r_{w,{\rm adv}}$
\begin{equation}
  \label{eq:27}
  r_{w, {\rm adv}} \simeq 50 \dot{m}_{w,1}
\end{equation}
Note that the advection radius is independent of the wind speed. Below $r_{w, {\rm adv}}$, photons are advected by the wind, so the photon temperature is controlled by adiabatic expansion. Since radiation pressure dominates, we have $P=aT^4/3\propto \rho^{4/3}\propto r^{-8/3}$, and we get
\begin{equation}
  \label{eq:30}
  T(r) = T_o (r/r_o)^{-2/3}
\end{equation}
Putting eq (\ref{eq:17}) and (\ref{eq:27}) into (\ref{eq:30}), we get the radiation temperature at the advection radius
\begin{equation}
  \label{eq:34}
T_{\rm adv}\equiv T(R_{w, {\rm adv}}) = 1.9\times 10^{5} r_{o,1}^{1/24} M_{\bullet,6}^{-1/4}
\dot{m}_{w,1}^{-5/12} \ \mathrm{K}
\end{equation}
Above $r_{w, {\rm adv}}$, photons are no longer advected by wind electrons, and the changing of color by Comptonization can be ignored. Therefore, the bolometric luminosity of the wind is
\begin{align}
  \label{eq:35}
      L_{w,{\rm bol}} &\simeq  \frac{4\pi R_{w, {\rm adv}}^2 aT_{\rm adv}^4v_w}{3} \\       
  L_{w,{\rm bol}} &\simeq 
  8.8\times10^{43} r_{o,1}^{-1/3} M_{\bullet,6} \dot{m}_{w,1}^{1/3} \ \mathrm{erg
  \ s^{-1}}
\end{align}
We can see the following: the wind luminosity can mildly exceed the Eddington luminosity (when $\dot{m}_w \gg1$); the wind is generally brighter than the other non-jet components, e.g., the disk and unbound debris \citep{2009MNRAS.400.2070S, 2010ApJ...714..155K}.

Many TDE candidates have very bright UV luminosities, which could come from the wind. For instance, TDE candidate PS1-10jh \citep{2012Natur.485..217G} has peak ($22$ d) luminosity $L\simeq 10^{45} {\rm erg/s}$ at $T_{BB}\sim 3\times10^4 {\rm K}$ ($T_{BB}$ has large uncertainties). By solving Eq.(\ref{eq:34}) and (\ref{eq:35}), we find $\dot{m}_w = 4.0\times10^2 r_{o,1}^{-1/8}$ and $M_{\bullet,6}=3.3r_{o,1}^{3/8}$.These values are roughly consistent with the accretion rate and BH mass expected in a TDE.

\subsection{External Inverse Compton Luminosity}
\label{sec:EIC_luminosity}
In this subsection, the EIC luminosities from the regions above and below the wind photosphere (cf. eq (\ref{eq:4})) are calculated, $L_{\rm IC}^{(1)}$ and $L_{\rm IC}^{(2)}$ respectively. For simplicity, we assume electrons have a single Lorentz factor $\gamma_e$ in the jet comoving frame. A powerlaw distribution is considered in next subsection, \S \ref{sec:EIC_spectrum}. We show that the EIC process above (below) the photosphere can boost the wind luminosity in Eq.(\ref{eq:35}) by a factor of $\sim \Gamma^2\gamma_e^2 \tau_{r} $ ($2\Gamma^2\gamma_e^2/\theta_j$). Therefore the observed X-ray luminosity ($3\times10^{47}$--$3\times10^{48}\ \mathrm{erg/s}$) is easily reached by boosting a wind luminosity ($10^{44}$--$10^{45}\ \mathrm{erg/s}$) by a factor of $10^2$--$10^3$.

\subsubsection{EIC above the photosphere}
Right above photosphere, $R\gae R_{w,{\rm tr}}$, the ERF is nearly isotropic. As we go farther from the photosphere, the flux from the photosphere decreases as $R^{-2}$ (inverse square law) and the back-scattered flux decreases as $R^{-3}$ (since $\tau_w\propto R^{-1}$). Since the photosphere flux is pointing parallel to the jet moving direction, the EIC emission comes mostly from the back-scattered flux
\begin{equation}
  \label{eq:47}
F_{\rm ex} (R)\simeq \frac{L_{w,{\rm bol}}}{4\pi R_{w,{\rm tr}}^2}
    \left( \frac{R_{w,{\rm tr}}}{R}
    \right)^{-3}
\end{equation}
where $L_{w,{\rm bol}}$ is the bolometric luminosity from the wind, eq (\ref{eq:35}). Since the flux in the back-scatter ERF drops off rapidly above the photosphere, the (isotropic) EIC luminosity from above the photosphere mostly comes from radii $R\sim R_{w,tr}$ 
\begin{align}
  \label{eq:42}
      L_{\rm IC}^{(1)} & \simeq
      \Gamma^2\gamma_e^2\left[F_{\rm ex}(R) \ \min{\left(\tau_{\rm trvs}(R),1 \right)}  2\pi R^2\theta_j
  \right]_{R_{w,{\rm tr}}} 
     \nonumber\\ 
     &\qquad\times\ \min{\left(\frac{4}{\theta_j^2}, 4\Gamma^2 \right)} \\
L_{\rm IC}^{(1)} &\simeq  \ \min{\left(
  1, \theta_j^2\Gamma^2 \right)} \Gamma^2\gamma_e^2
\tau_{r}(R_{w,{\rm tr}}) L_{w,{\rm bol}}
\end{align}
where $\tau_{r}$ is the jet optical depth in the radial direction, see eq (\ref{eq:9}), and in the second approximation we have used $R_{w,{\rm tr}}>R_{j,{\rm self}}$ (usually true). We can see	 that if $\theta_j\gae 1/\Gamma$, the EIC process above the photosphere boosts the wind luminosity by a factor of $\sim\Gamma^2\gamma_e^2 \tau_{r}(R_{w,{\rm tr}})$.

\subsubsection{EIC below the photosphere}
Below the photosphere, $R\lae R_{w,{\rm tr}}$, far away from the jet funnel, the radiation energy density $U$ is a function of radius $R$. Below $R_{w,{\rm adv}}$, radiation suffers from adiabatic expansion $U \propto R^{-8/3}$; above $R_{w,{\rm adv}}$,  diffusion is in control and the diffusive flux is $F_{\rm dif} \simeq Uc/3\tau_{w} \propto R^{-2}$. Near the jet funnel, the diffusive flux entering the jet funnel is
\begin{equation}
  \label{eq:12}
    F_{\rm dif}(R) \simeq \frac{U(R)c}{3\tau_w(R)}
 \propto
 	\left\{
    \begin{array}{ll}
      R^{-5/3} & \mbox{if } R < R_{w,{\rm adv}}\\
 R^{-2} & \mbox{if } R_{w,{\rm adv}}<R<R_{w,{\rm tr}}
    \end{array}\right.
\end{equation}
At radius $R<R_{j,{\rm self}}$, the flux entering the funnel is immediately scattered by the jet. At radii $R_{j,{\rm self}}<R<R_{w,{\rm tr}}$, photons can penetrate the jet in the transverse direction, so the radiation field in the funnel tends to isotropize and approach the same energy density as in the surrounding wind. The number of photons escaping along the radial direction of the funnel are negligible. The only process that prevents the isotropization of the UV photons is the removal of photons by jet scattering. Therefore, we expect the radiation field to isotropize when the jet optical depth $\tau_{\rm trvs}$ is too small to remove a large fraction of photons away from the funnel---i.e., when the removing rate equals the injecting rate
\begin{equation}
  \label{eq:45}
  \tau_{\rm trvs}Uc = \frac{Uc}{3\tau_w} \mbox{, i.e. }
  \tau_{\rm trvs} \tau_w = \frac{1}{3}
\end{equation}
Note that we use $\tau_{\rm trvs}$ instead of $\tau_{r}$ because the isotropization happens locally (from $R$ to $R+dR$) in the transverse direction. From Eq.(\ref{eq:45}), we get $r_{w,{\rm iso}}$, the (dimensionless) ``isotropization radius'' \begin{equation}
  \label{eq:41}
r_{w,{\rm iso}} \simeq 96 M_{\bullet,6}^{-1/2}\dot{m}_{w,1}^{1/2}
      r_{o,1}^{1/4} 
  \left( \frac{L_{j,48} \theta_{j,-1}}{\Gamma_1} \right)^{1/2}
\end{equation}
Above $r_{w,{\rm iso}}$, the energy density of ERF will reach the same value inside of the funnel as the wind, $U$. Comparing eq (\ref{eq:41}) with eq (\ref{eq:27}), we can see that the difference between $r_{w,{\rm adv}}$ and $r_{w,{\rm iso}}$ is usually small.

Therefore, the (isotropic) EIC luminosity from below $R_{w,{\rm tr}}$ is
\begin{align}
L_{\rm IC}^{(2)} & \simeq  
 \Gamma^2\gamma_e^2  \int_{R_{\rm min}}^{R_{w,{\rm tr}}}
  \min{\left( \tau_{\rm trvs}Uc,
  \frac{Uc}{3\tau_w} \right)} 2\pi R\theta_j dR\nonumber\\ 
  &\qquad\times \ \min{\left(
         \frac{4}{\theta_j^2}, 4\Gamma^2 \right)}
        \\
\label{eq:46}  
L_{\rm IC}^{(2)} &\simeq 
  L_{w,{\rm bol}} \frac{2\gamma_e^2\Gamma^2}{\theta_j}\min{\left[ 1, \left(
    \frac{R_{w,{\rm iso}}}{R_{w,{\rm adv}}} \right)^{1/3} \right]} \nonumber\\
    &\qquad\times \min{\left(1, \theta_j^2\Gamma^2\right)}
\end{align}
where the the lower limit of the integration, $R_{\rm min}$, is the radius where the jet accelerated to high Lorentz factor $\Gamma$. Below the wind launching radius $R_o\simeq 2R_p$, the ERF possibly comes directly from the inner disk, which is theoretically very uncertain\footnote{See the discussion about super-Eddington accretion  with mass loss in \citet{1973A&A....24..337S, 2007MNRAS.377.1187P, 2011MNRAS.413.1623D, 2012MNRAS.420.2912B}.}. Fortunately,  the contribution from the radii near $R_o$ is small, because most of the EIC luminosity comes from the region $R\simeq \mbox{min}(R_{w,{\rm iso}}, R_{w,{\rm adv}})\gg R_o$. We can see that the EIC process below the photosphere boosts the wind luminosity by a factor of $\sim 2\Gamma^2\gamma_e^2/\theta_j$.

\subsection{External Inverse Compton Spectrum}
\label{sec:EIC_spectrum}
As shown in subsection \ref{sec:EIC_luminosity}, EIC emission could come from above and below the photosphere, with EIC luminosities given by $L_{\rm IC}^{(1)}$ (eq \ref{eq:42}) and $L_{\rm IC}^{(2)}$ (eq \ref{eq:46}) respectively. Usually, $L_{\rm IC}^{(1)}\ll L_{\rm IC}^{(2)}$. Therefore, we consider the spectrum from below the photosphere ($R<R_{w,{\rm tr}}$) first and see if it can reproduce the observed X-ray power law of Sw J1644+57.

Since most of the EIC  luminosity comes from the region $R\sim R_{w,{\rm iso}}\sim R_{w,{\rm adv}}$ (for simplicity, we ignore the difference between $R_{w,{\rm iso}}$ and $R_{w,{\rm adv}}$ hereafter), the $\nu f_\nu$ peak energy of the EIC spectrum is
\begin{equation}
  \label{eq:6}
  h\nu_p \simeq \Gamma^2\gamma_e^2 3kT_{\rm adv} \simeq 5 \Gamma_1^2 
  \gamma_e^2 r_{o,1}^{1/24} M_{\bullet, 6}^{-1/4} \dot{m}_{w,1}^{-5/12} \ \mathrm{keV}
\end{equation}
From eq (\ref{eq:30}) and (\ref{eq:46}), we can see that the EIC emission from below the advection radius ($R<R_{w,{\rm adv}}$) has a power-law shape
\begin{equation}
  \label{eq:14}
  \frac{dL_{\rm IC}}{dT}=\frac{dL_{\rm IC}}{dR} \frac{dR}{dT}\propto \frac{R
    U(R)}{\tau_{w}(R)} R^{5/3}\propto T^{-3/2}
\end{equation}
Therefore, the EIC spectrum above the peak energy $\nu_p$ is $F_{\nu}\propto \nu^{-3/2}$, which is too soft to explain the observations ($F_{\nu}\propto \nu^{-0.8}$ or $\propto \nu^{-0.4}$ at late time). Around $\nu_{p}$, there's a big bump coming from the EIC emission at radii $R>R_{w,{\rm adv}}$. Below $\nu_{p}$; we expect a standard inverse Compton low energy spectrum $f_\nu \propto \nu$ (cf. \S\ref{sec:IC}, \citealt{Blumenthal1970}) Simply tuning $\Gamma$ or $\gamma_e$ will not work because changing these parameters only changes the position of $\nu_p$ but not the the high-energy power-law index.

A possible solution is the case where electrons in the jet have a power-law distribution $dN_e/d\gamma_e \propto \gamma_e^{-p}$ ($\gamma_{e, {\rm min}} < \gamma_e < \gamma_{e, {\rm max}}$). If so, we expect the high-energy EIC spectrum to be $f_\nu\propto \nu^{-(p-1)/2}$. An electron population with $p=2.6$ will reproduce the observed power-law\footnote{At later time, the observed spectrum  becomes slightly harder ($f_\nu\propto\nu^{-0.4}$). From eq (\ref{eq:34}), we can see that $T_{\rm adv}$ increases with time, so the EIC spectrum in the XRT band ($0.3-10\ \mathrm{keV}$) will get harder at later time.} $F_{\nu}\propto \nu^{-0.8}$. The range of electrons' Lorentz factors must satisfy $\gamma_{e,{\rm max}}/\gamma_{e,{\rm min}} \geq (10\ \mathrm{keV}/0.3 \ \mathrm{keV})^{1/2}\simeq 5.8$, or $ (150\ \mathrm{keV}/0.3 \ \mathrm{keV})^{1/2}\sim 20$ (if including the BAT data). Convolving eq (\ref{eq:46}) with the electrons' Lorentz factor distribution, we find $L_X$, the isotropic EIC luminosity within the XRT band ($0.3-10\ \mathrm{keV}$),
\begin{equation}
  \label{eq:18}
  L_{0.3-10} \simeq \frac{8\Gamma^2 \gamma_{e,{\rm min}}^2}{\theta_j}
      \left(\frac{\gamma_{e,0.3\mathrm{keV}}}{\gamma_{e,{\rm min}}}\right)^{3-p} L_{w,{\rm bol}}
\end{equation}
The approximation is accurate within a factor of 20\% for $2<p<3$. In eq (\ref{eq:18}), $\gamma_{e,0.3\mathrm{keV}}$ is the electron's Lorentz factor corresponding to scattered photons' energy 0.3 keV.

To reproduce the XRT observations, we must match both the luminosity in the $0.3-10\ \mathrm{keV}$ band, and the power-law spectrum. We have two equations:
\begin{equation}
  \label{eq:16a}
 \Gamma^2\gamma_{e, {\rm min}}^2 3kT_{\rm adv} \leq \Gamma^2\gamma_{e,
      0.3\mathrm{keV}}^2 3kT_{\rm adv}=   0.3 (1+z) \ \mathrm{keV} 
\end{equation}
\begin{equation}
L_X\simeq
\frac{8\Gamma^2 \gamma_{e,{\rm min}}^2}{\theta_j}
      \left(\frac{\gamma_{e,0.3\mathrm{keV}}}{\gamma_{e,{\rm min}}}\right)^{3-p}
      L_{w,{\rm bol}}
\end{equation}
Using $z = 0.35$, $p = 2.6$, $T_{\rm adv}$ from eq (\ref{eq:34}), $L_{w,{\rm bol}}$ from eq (\ref{eq:35}) and $L_{X} =
10^{47}L_{X,47} \ \mathrm{erg \ s^{-1}}$,
we find 
\begin{align}
  \label{eq:16b}
 \Gamma\gamma_{e,{\rm min}} & \lae  3\  r_{o,1}^{-1/48} M_{\bullet, 6}^{1/8}
    \dot{m}_{w, 1}^{5/24} \\
\Gamma \gamma_{e, {\rm min}} &\simeq  4 \ \theta_j^{5/8} L_{X,47}^{5/8}
r_{o,1}^{5/24} M_{\bullet,6}^{-5/8} \dot{m}_{w,1}^{-5/32}
\end{align}
If we use $\theta_j\simeq 1/\Gamma$, the second inequality restricts the jet Lorentz factor to a very small value $\Gamma\lae 2$, which is inconsistent with the constraints from variability time scale $\delta t\simeq 10^2\ \mathrm{s}$. 

If the EIC emission comes from above the photosphere ($L_{\rm IC}^{(1)}$ in eq \ref{eq:42}), we substitute $\theta_j$ in the second inequality above by $2/\tau_{r}(R_{w,{\rm tr}})$, i.e.
\begin{align}
  \label{eq:23}
 \Gamma\gamma_{e, {\rm min}} &\lae 3\ r_{o,1}^{-1/48} M_{\bullet,6}^{1/8}
    \dot{m}_{w, 1}^{5/24} \\
 \Gamma \gamma_{e, {\rm min}} & \simeq  6 \ \tau_{r}^{-5/8}(R_{w,{\rm tr}})
L_{X,47}^{5/8} r_{o,1}^{5/24} M_{\bullet,6}^{-5/8} \dot{m}_{w,1}^{-5/32}
\end{align}
Since $\tau_{r}(R_{w,{\rm tr}})$ can range from $0.1-1$, the EIC emission from above the photosphere gives less stringent constraints on the Lorentz factors than from below the photosphere. Recall that the wind mass loss rate is $\dot{m}_w=f_{\rm out}\dot{m}_{\rm fb}$. In the first few weeks $\dot{m}_{\rm fb}\sim 10^2$, so the EIC model is marginally consistent with the observations if: (1) the EIC emission comes from above the photosphere, (2) the wind mass loss has high efficiency $f_{\rm out}\simeq 1$, (3) the jet Lorentz factor is not too high $\Gamma\lae 5$, and (4) electrons' minimum Lorentz factor $\gamma_{e,min}\simeq 1$. later time, the X-ray light-curve drops roughly as $t^{-5/3}$, which strongly suggests that the X-ray luminosity tracks the jet kinetic power and hence the jet radiation efficiency is roughly constant (likely $\sim 10\%$). From equation (\ref{eq:EIC_cool_ratio}), substituting $m_p$ with $\gamma_e m_e$, we can see that all electrons are in the fast cooling regime even at late time when $L_{\rm disk} \sim 10^{43}\ {\rm erg/s}$, so they must be reheated continuously. Therefore, the EIC model requires that the unknown (but non-Coulomb) reheating mechanism pass a roughly constant fraction $\xi\sim 10\%$ of the jet power into electrons and then quickly into radiation, i.e.

\begin{equation}
L_{\rm EIC} = \xi L_j = U_{\rm ERF} \sigma_T c \Gamma^2 \int_{\gamma_{e,{\rm min}}}^{\gamma_{e,{\rm max}}}\gamma_e^2 \frac{dN_e}{d\gamma_e} d \gamma_e
\end{equation}
or
\begin{equation}
\xi = U_{\rm ERF} R \frac{\Gamma \sigma_T}{m_p c^2} \frac{p-1}{3-p} \frac{\gamma_{e,{\rm max}}^{3-p} - \gamma_{e, {\rm min}}^{3-p}}{\gamma_{e,{\rm min}}^{1-p} - \gamma_{e,{\rm max}}^{1-p}}
\end{equation}
where $R$ is the radius where most EIC luminosity is produced and $U_{\rm ERF}(R)$ is the energy density of the external radiation field at that radius. We can see that $\xi$ can be a constant only when the product $R U_{\rm ERF} \propto L_{\rm ERF}/R$ stays roughly constant with time. Otherwise, the EIC model needs fine tuning on parameters such as $\Gamma$ and $\gamma_{e,max}$. In the first few weeks, we estimate $L_{\rm ERF} \sim 10^{45}\ {\rm erg/s}$ and $R\sim 10^{15}\ {\rm cm}$, and after a few hundred days, we have $L_{\rm ERF}\sim 10^{43}\ {\rm erg/s}$ and $R\sim 10^{13}\ {\rm cm}$. However, the quantity $L_{ERF}/R$ is highly uncertain due to the lack of UV data and if it fluctuates by a factor of $>3$ (which seems likely to the authors), the EIC model requires fine tuning.

\section{Photospheric model of TDE Sw J1644+57}
\label{sec:Photospheric}

Another mechanism studied to explain non-thermal emission from relativistic jets, in the context of GRBs, is the photospheric process \citep{1994MNRAS.270..480T, 1999ApJ...511L..93G, 2000ApJ...530..292M, 2005ApJ...628..847R, 2007ApJ...666.1012T, 2008ApJ...682..463P, 2008A&A...480..305G}. In this section, we present results from Monte Carlo (MC) simulations using a recent code presented in \citet{Santana2015} to determine if the photospheric model can explain the observed $f_\nu\propto\nu^{-0.8}$ X-ray spectrum  of Sw J1644+57. We first describe the photospheric process and the conditions that need to be satisfied to apply the photospheric process to Sw J1644+57. Then, we briefly discuss the algorithm used and the parameters we consider in our MC photospheric simulations. Lastly, we discuss the simulation results and compare them to the Sw J1644+57 observations. 

\subsection{Spectrum for the Photospheric Process}

The photospheric process involves photons undergoing multiple scatterings (Comptonization) with electrons that are heated close to the photosphere, where the medium is still optically thick. Since the emission takes place below the photosphere, the photospheric process is typically invoked as the hot electrons reprocessing a Blackbody (BB) spectrum. The effect of Comptonizing a BB spectrum of photons is to broaden the BB spectrum, adding a non thermal tail above the peak. The photospheric process is predicted to produce a spectrum below the BB peak-energy ranging from $f_{\nu} \propto \nu^{2}$ to 
$f_{\nu} \propto \nu^{0}$ \citep{2013ApJ...764..143V, 2013MNRAS.428.2430L, 2014ApJ...785..112D} and a typical spectrum above the BB peak-energy $f_{\nu} \propto \nu^{-1}$ \citep{Lazzati2010}. For the photospheric process to work for Sw J1644+57, the high energy part of the spectrum will have to match the observed spectral index in the XRT/BAT, $\beta\approx0.8$.

The basic picture of the photospheric process is as follows. When the optical depth of the outflow is $\sim 2-20$, a dissipation event is assumed to occur, which accelerates electrons to mildly-relativistic or relativistic speeds. In the photospheric process, the temperature of the photons is taken to be much smaller than the temperature of the electrons. Therefore, the photons continue to gain energy from electron scattering until the energy of the photons reaches the temperature of the electrons or until the photons escape the photosphere (i.e. the photons stop scattering off of electrons). A power-law spectrum is expected to be produced above the initial BB peak since only a fraction $f$ of the photons get scattered once by a hot electron to higher energies, and therefore a fraction $f^{2}$ of the photons get scattered twice by a hot electron to higher energies, and so on \citep{Lazzati2010,2013LNP...873.....G}.

In the next subsection, we determine if there is a parameter space where the jet is optically thick so that we can apply the photospheric process.\

\subsection{Photospheric Radius for Sw J1644+57}
\label{gam_ratio_phot_elec}

The optical depth to Thomson scattering if the jet is baryon-dominated without a significant number of pairs is given as $\tau_{\rm jet}$ in eq (\ref{eq:tau_jet}). From the equation for $\tau$, we can determine the location of the photospheric radius ($R_{\mathrm{ph}}$), i.e. the location where the medium becomes optically thin and the photon spectrum is no longer significantly modified by electron scatterings. Setting $\tau = 1$ in equation \ref{eq:tau_jet} and solving for $R$, we find 
\begin{equation}
R_{\mathrm{ph}} = \frac{L_j \sigma_{T}}{4 \pi m_{p} c^{3} \Gamma^{3}} = 
(1.2 \times 10^{12} \mathrm{ cm}) L_{j,48} \Gamma_{1} ^{-3}.
\label{rphot_eq_phot}
\end{equation}
In order to ensure that there is a parameter space where the medium is optically thick, the condition 
$R_{S} < R_{\mathrm{ph}}$ must be satisfied, where $R_{S}$ is the Schwarzschild radius, $R_{S} = (3 \times 10^{11} \mathrm{cm}) M_{\bullet,6}$, where $M_{6}$ is the mass of the SMBH in units of $10^{6} M_{\odot}$. For $\Gamma=10$, $M_{\bullet,6}= 4.3$, $L_{j,48} =  1$, and (corresponding to a $\sim$10\% efficiency), $R_{\mathrm{ph}}$ is roughly equal to $R_{S}$. Therefore, the photospheric processes will only work for modest $\Gamma$, that is, to ensure than $R_{S} < R_{\mathrm{ph}}$; $2 \le \Gamma \lae 10$. 

\subsection{Brief Description of the MC Photospheric Simulations}

In \citet{Santana2015}, the photospheric code is described in great detail, but we briefly describe the code in this section. In the photospheric process, the electrons are initially hotter than the photons. Thermal photons scatter off of hot electrons to gain energy and populate a non-thermal power law. To simulate this process, we use Monte Carlo (MC) methods to create an array of electrons and an array of photons, each drawn from their respective parent distribution. We then use MC methods to allow the two species interact with one another via Compton scattering. The electrons are assumed to be uniform throughout the jet. For Sw J1644+57, we take the electrons to be part of a Maxwell-Boltzmann distribution with electron temperature $T'_e$ in the co-moving frame. \citet{Santana2015} showed that while the energy carried by the electrons can affect the output spectra greatly, the choice of electron distribution has little effect. Our findings would be similar for other electron distributions.

\begin{figure*}
\begin{center}
$\begin{array}{cc}
\includegraphics[width=0.48\textwidth]{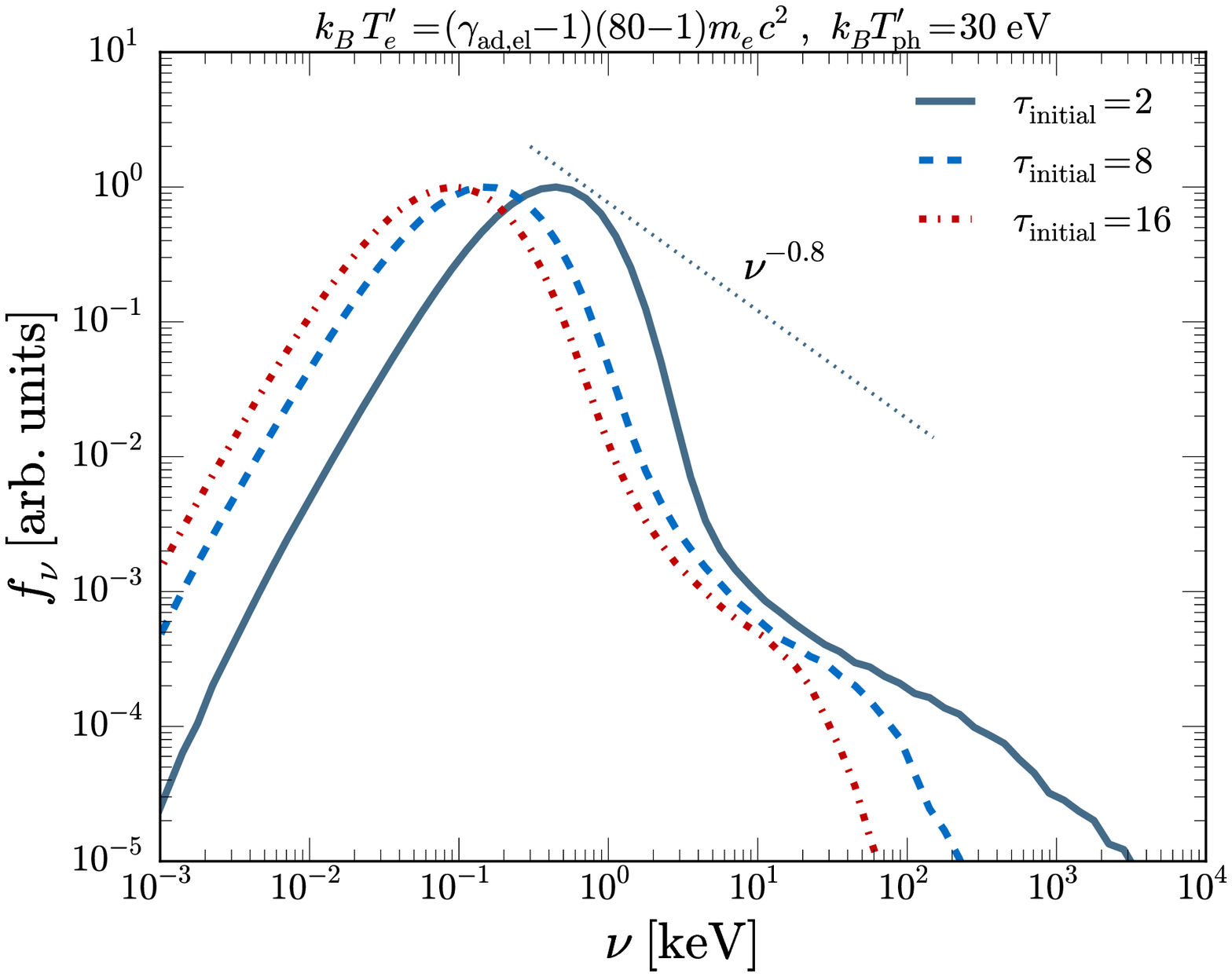} &
\includegraphics[width=0.48\textwidth]{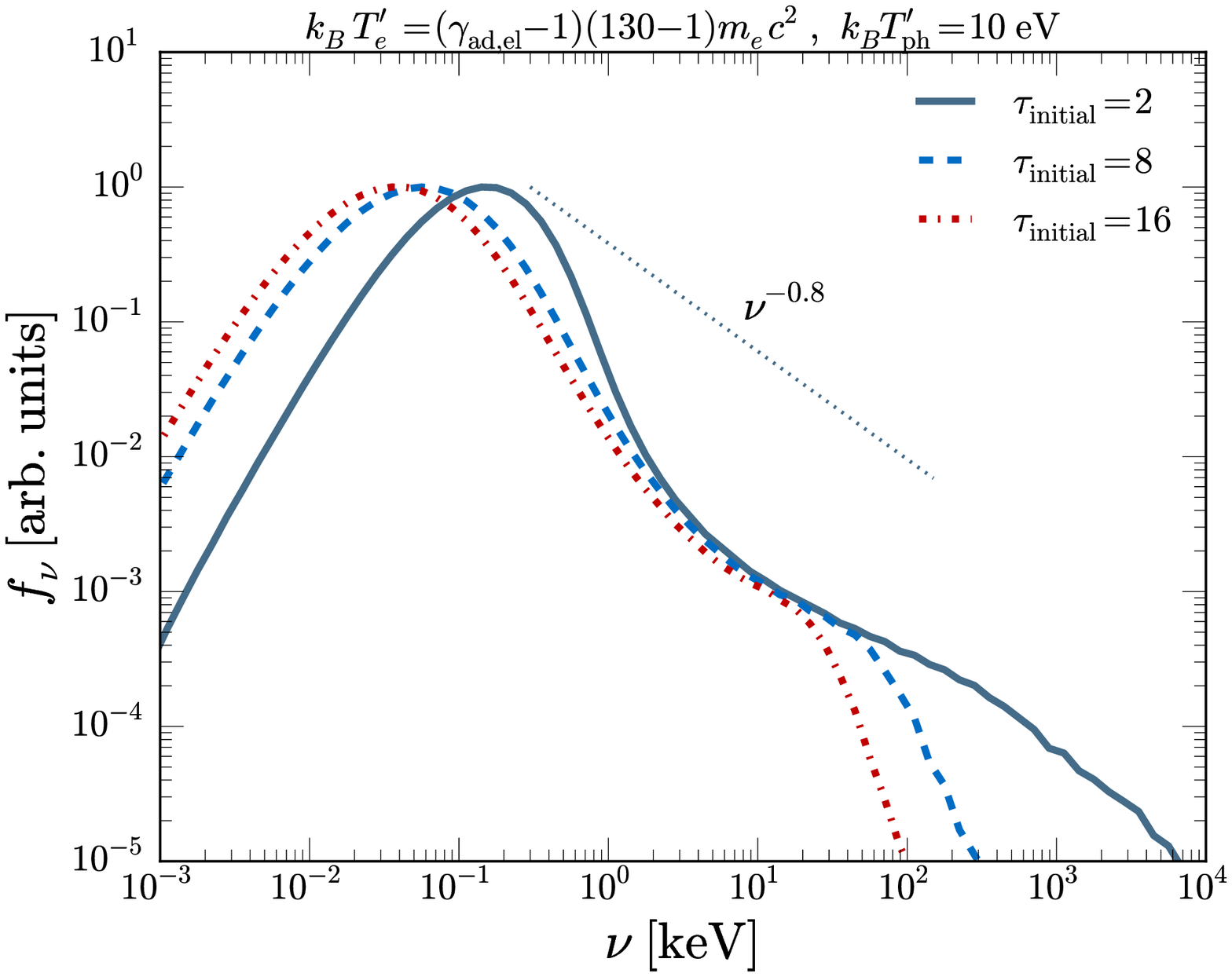}
\end{array}$
\end{center}
\caption{Output spectrum for MC photospheric simulations for Maxwell Boltzmann (MB) electrons, $\tau_{\mathrm{initial}} = $ 2, 8, 16, and  $\Gamma=5$. The lines correspond to the photon spectrum at the end of the simulations in the observer frame. A power-law with $\beta = 0.8$ is shown for reference. In Sw J1644+57, the photospheric process is unable to produce a hard power-law above the peak without a low efficiency or multiple reheating events because the electrons cool too quickly.}
\label{fig:photospheric_spec}
\end{figure*}

The photons are initialized as a Blackbody (BB) distribution with co-moving temperature $T'_\gamma$. The photons are initialized at an optical depth $\tau_{\rm initial}$, with an equivalent radius $R_{\rm initial}$. The photons' positions and momenta are tracked. For each photon, we draw a random `travel length' that depends on the mean free path at the current location of the photon in the jet. This length is the distance that each photon will travel before it interacts with an electron. We take the photon with the smallest travel length and draw an electron randomly from the electron array to scatter this photon. Before we perform this scattering, we calculate the radius where the scattering takes place to take into account the energy the photon and electrons adibatically lose in traveling from the initial radius to the radius where the scattering takes place. Next, we scatter the photon off the randomly drawn electron. We calculate the outgoing momenta and energy of the electron and photon, and then draw a new travel length for the photon. If this length would allow it to travel outside of the photosphere, we collect the photon as part of the observed spectrum, and return the electron to its array. Otherwise, we return the electron and the photon back to their arrays with the new energies, momenta, and travel length. We continue to loop over this process, each time drawing the photon with the new smallest travel length. We stop looping when one third of the initialized photons have escaped the photosphere. By collecting a third of the photons, we are measuring a time-averaged spectrum.

To accurately model the photospheric process, the MC simulation needs the correct ratio of the jet's energy carried by electrons compared to the photons. This is done by fixing the ratio of photons to electrons, $N_\gamma/N_E$. \citep{Lazzati2010,2015ApJ...802..132C}.  \citealt{Santana2015} show that the photon to electron ratio is given by
\begin{equation}
\frac{N_{\gamma}}{N_{e}} = (5 \times 10^{7}) \left( \frac{\eta}{1-\eta} \right) \left( \frac{\Gamma}{5} \right)
\left( \frac{E_{\mathrm{pk}}}{\mbox{ 0.1 keV}} \right)^{-1} ,
\label{photon_electron_ratio_eq}
\end{equation}
where $E_{\mathrm{pk}}$ is the peak energy of the spectrum for Sw J1644+57, in the observer frame, and $\eta$ is the efficiency of the production of X-rays (i.e. $\eta L_j = L_X$). The value of $\Gamma$ was normalized to 5 since with equation \ref{rphot_eq_phot} we determined that $2 \le \Gamma \lae 10$. $E_{\mathrm{pk}}$ was normalized to 0.1 keV. Since Sw J1644+57 doesn't show a peak in the 0.3-10keV XRT band, the peak energy must be below 0.3 keV. With a 10\% efficiency in the jet, the photon to electron ratio is  $N_{\gamma}/N_{e} \sim 6 \times 10^{6}$. We are unable to model a ratio of $N_{\gamma}/N_{e} = 6 \times 10^{6}$ as it is too computationally expensive. Instead, we will consider $N_{\gamma}/N_{e} = 2 \times 10^{6}$. Increasing $N_{\gamma}/N_e$ will produce a softer power-law tail for the photospheric spectrum \citep{Santana2015}. Therefore, if the photospheric process is unable to match the observed spectrum of Sw J1644+57 with a slightly underestimated $N_\gamma/N_e$, it will not be able to do so with the real value.

In the simulations, we set the number of electrons, $N_{e}$, to 100. To preserve the correct $N_\gamma/N_e$ ratio, we set $N_\gamma$ to $2\times10^8$. Previous works typically considered $N_{e} = 10^{3}$ \citep{Lazzati2010, 2015ApJ...802..132C}. To determine if $N_{e} = 10^{2}$ electrons are enough for a MC photospheric simulation, we performed a few simulations with $N_{e} = 10^{3}$ and $N_\gamma = 2\times10^9$ found the results to be identical to the  $N_{e} = 10^{2}$ simulation results. Therefore, $N_{e} = 10^{2}$ is enough electrons to capture the important physics of the photospheric process.

The energy carried by the electrons, $\gamma_{e} ^{\prime}$ or $T_{e} ^{\prime}$, can drastically affect the observed spectrum in the photospheric process. $\gamma_{e} ^{\prime}$ is a very important parameter since it determines the available energy electrons have to transfer to photons. In order to avoid synchrotron cooling from the magnetic field that is expected to be present in the jet, the synchrotron emission must be self-absorbed. The largest $\gamma_{e}^{\prime}$ that can be considered is found by setting the optical depth for synchrotron self-absorption equal to 1. For a Maxwell-Boltzmann distribution distribution of electrons, the synchrotron self-absorption optical depth $\tau_{\mathrm{syn}}^{\mathrm{MB}}$ is given by \citep[][]{RybickiLightman,Lazzati2010}
\begin{equation}
\tau_{\mathrm{syn}} ^{\mathrm{MB}} = \frac{10^{10}}{(\gamma_{e} ^{\prime} ) ^{5} 
\epsilon_{B} ^{1/2} (E _{ \gamma} ^{\prime} / \mbox{10 eV}) ^{2}} .
\label{self_abs_phot_sect}
\end{equation}
In this equation, $\epsilon_{B} = U'_{B}/U'_{\mathrm{rad}}$, where $U'_{B}$ ($U'_{\mathrm{rad}}$) is the energy density in the magnetic field (radiation). Setting $\tau_{\mathrm{syn}}^{\mathrm{MB}} = 1$, for $\epsilon_{B} \sim 0.1$ (magnetic field sub-dominant to radiation), for the  photon temperatures we consider, $E_{\gamma}^{\prime} \sim $ 10 eV, and 30 eV the upper limits we find are $\gamma_{e}^{\prime} \sim $ 125, and 80 respectively. We consider the largest value allowed for $\gamma_{e}^{\prime}$ for each photon temperature we consider. Since $k_{B} T_{e}^{\prime}$ measures the kinetic energy of electrons, the value of $T_{e}^{\prime}$ corresponding to a $\gamma_{e} ^{\prime}$ value is found with the expression $k_{B} T_{e}^{\prime} = (\gamma_{\mathrm{ad, el}} - 1)  (\gamma_{e}^{\prime} - 1)$, where $\gamma_{\mathrm{ad, el}} = (4 \gamma_{e}^{\prime} + 1)/(3 \gamma_{e}^{\prime})$ is the electron adiabatic index.

\subsection{MC Photospheric Simulation Results}

In Figure \ref{fig:photospheric_spec}, we show the results from our simulations. We assume $\Gamma=5$ and set total luminosity in the jet, $L_j$, to $10^{48}$ erg/s. We consider two different BB temperatures for the photons, $T_{\gamma}^{\prime} = $ 30 eV in the left panel and 10 eV in the right. These co-moving temperatures correspond to an observed peaks at 0.15 keV and 0.05 keV, so any powerlaw that extends for a order of magnitude or so above the peak will be in the Swift XRT energy range. We give the electrons as high of $\gamma'_e$ as possible, still requiring that they are synchrotron self-absorbed. For each $T'_\gamma$, we initialize the photons at 3 different optical depths, $\tau_{\mathrm{initial}} = $ 2, 8, 16.

In both of the panels, the photon spectra do not show the development of a hard power-law above the peak-energy. For larger optical depths, the additional scatterings allow for more photons to be upscattered to higher energies, making it easier to produce a power-law spectrum above the peak-energy, but in all the cases, the power-law that forms is softer than $\nu^{-0.8}$. In Figure \ref{fig:photospheric_spec}, the $\tau_{\mathrm{initial}} = $ 8, 16 simulations with lower $k_{B} T_{\gamma}^{\prime}$ display a harder spectrum above the peak-energy for the following two reasons: 1. lower energy photons cool the electrons more slowly, allowing for more photons to be upscattered to higher energies. 2. For the lower $k_{B} T_{\gamma} ^{\prime}$ simulations, the electrons have more energy to transfer to the photons. This is due to the self-absorption condition (Equation \ref{self_abs_phot_sect}); we can consider larger $\gamma_{e}^{\prime}$ for lower $k_{B} T_{\gamma}^{\prime}$. Finally, we note that the peak-energy for the $\tau_{\mathrm{initial}} = $ 8, 16 simulations is lower than the peak-energy of the $\tau_{\mathrm{initial}} = $ 2 simulations due to the increased adiabatic cooling of the photons. 

We also note that the simulation results we presented were for the scenario where electrons were only accelerated once at the start of the simulation. It is also possible that the electrons are re-heated by either magnetic reconnection or multiple shocks while the jet is still optically thick. If the electrons have more energy, then more photons can be upscattered to higher energies and thus a shallower spectrum can be produced above the peak-energy. We can estimate the number of reheating events in a similar manner as \citet{Santana2015}
\begin{equation}
N_{\mathrm{reheat}} \sim \frac{N_\gamma}{N_e}\frac{E'_\gamma}{m_e c^2}\left\{\ln{\left[\frac{(\gamma_e-1)}{(\gamma_e+1)}\right]}+1.5\right\}^{-1}\sim 10^3\frac{\eta}{1-\eta}
\end{equation}
Where we used eq (\ref{photon_electron_ratio_eq}) to derive the right hand side of the equation above. For an efficiency of 10\%, we find that there would need to be $\sim 100$ heating events to produce a hard spectrum in agreement with the observed $f_{\nu} \propto \nu^{-0.8}$ spectrum. However, even if we were to consider electron re-heating or a lower efficiency of $\sim1\%$, the photospheric process suffers from another problem. At late times, the X-ray luminosity for Sw J1644+57 drops by a factor $\sim 100$. From Equation \ref{eq:tau_jet}, a decreases in $L$ by a factor of 100 would decrease $\tau$ by a factor of 100. Therefore, at late time, the jet is expected to be optically thin and the photospheric process can no longer be applied to Sw J1644+57. 

We conclude that the photospheric process is very unlikely to explain the Sw J1644+57 observations since it has a difficult time producing the observed $f_{\nu} \propto \nu^{-0.8}$ spectrum and it cannot be applied at late times since the jet is expected to be optically thin.
\section{Poynting-Dominated jet}
\label{sec:Poynting}
The idea that the relativistic jets in AGN extract energy from the SMBH through magnetic fields through some sort of Blandford-Znajek \citep{1977MNRAS.179..433B} process has observational and theoretical evidence \citep[e.g.,][]{1984RvMP...56..255B,2004ApJ...605..656V,2011MNRAS.418L..79T, 2014Natur.515..376G}. The jets in tidal disruptions such as Sw J1644+57 may be powered by magnetic flux, but the magnetic flux from the star that is disrupted is not large enough to power a TDE without some sort of dynamo that is capable of generating a large scale magnetic field \citep{Bloom2011}. \citet{2014MNRAS.445.3919K} have suggested a solution to this problem; the stellar debris may bring in the magnetic flux from a fossil accretion disk as it falls back onto the disk. The possibility that Sw J1644+57 is magnetically dominated was considered by \citet{Burrows2011, 2014MNRAS.437.2744T,2015MNRAS.453..157P}. In this section we model the consequences of a Poynting-dominated jet on the observed radiation of Sw J1644+57.

How the magnetic energy is dissipated in astrophysical systems through reconnection is not well understood. Instead of modeling the details of the reconnection process, we use the general considerations for any reconnection process derived through basic global properties of the magnetic field. We provide these derivations in \citet{KumarCrumley2015}.

As we have shown in the previous sections, fast cooling of the keV producing electrons is nearly unavoidable in Sw J1644+57; this is because the cooling time is very short when compared to the dynamical time. This short cooling time forces the source injecting hot electrons to do so rapidly. If the electrons are not reheated, excess flux can occur at lower wavelengths. Magnetic reconnection models can avoid this problem in two ways. First, the acceleration process can happen throughout a large domain, and the electrons can spend a significant fraction of time in these acceleration regions. In acceleration regions, the electrons will not cool significantly; the acceleration process continues until the rate at which the electrons lose energy is greater than the rate at which they gain energy---at this point, the electrons will stay at this Lorentz factor, $\gamma_{\rm max}$. Secondly, there are multiple reconnection regions that are radiating all at once. After leaving a reconnection region, the electron will cool rapidly, as before. However since there are many acceleration sites throughout the jet, electrons may enter into another reconnection site and be reaccelerated.

When the jet is magnetically dominated, the jet's luminosity is roughly equal to the luminosity carried by the magnetic field, so magnetic field can be written in terms of the isotropic equivalent luminosity of the jet, $L$, 
\begin{equation}\label{eq:mag_B}
 B'_0 = \frac{(L/c)^{1/2}}{\Gamma R} = (580\ \mathrm{G})\ \frac{L_{48}^{1/2}}{\Gamma_1 R_{15}};
\end{equation}
If electrons in the particle acceleration region are producing keV synchrotron photons, the typical electron Lorentz factor must be
\begin{equation}\label{LFi}
   \gamma_i \approx \left[ \frac{2\pi m_e \nu R c^{3/2}}{q L^{1/2}(1+z)^{-1}}
          \right]^{1/2} = 4\times10^3 \frac{[R_{15} \nu_\mathrm{kev}(1+z)]^{1/2}}
        {L_{48}^{1/4}}.
\end{equation}
The total number of electrons radiating at 1 keV required to produce an observed flux $f_\nu$ is
\begin{equation}
   N_e \approx 1.2	\times10^{49} f_{\nu,\mathrm{mJy}} L_{48}^{-1/2} R_{15} d_{L,28}^2 
         (1+z)^{-1},
   \label{Ne}
\end{equation}
and the corresponding luminosity carried by these electrons is 
\begin{align}
    L_e & \approx  \frac{N_e m_e c^3 \gamma_e \Gamma}{(R/\Gamma^2)} \nonumber \\
        & = (1.2\times10^{42}\ \mathrm{erg\, s^{-1}}) \frac{\Gamma^3 f_{\nu,mJy} 
     R_{15}^{1/2} d_{L,28}^2 \nu_{\rm keV}^{1/2}}{L_{48}^{3/4}(1+z)^{1/2}}.
\end{align}
The electrons get all their energy from the magnetic fields in Poynting dominated models. Therefore, the energy carried by the electrons cannot exceed the energy in magnetic fields prior to reconnection, placing the following constraint on the bulk Lorentz factor of the jet:
\begin{equation}
    L_e/L \lae 1 \quad\Longrightarrow\quad \Gamma \lae 90\; \frac{ L_{48}^{7/12} 
     (1+z)^{1/6}}{f_{\nu,mJy}^{1/3} R_{15}^{1/6} d_{L,28}^{2/3} 
     \nu_{keV}^{1/6} }.
\end{equation}
The reason for the approximate inequality sign in the above equation is because magnetic fields of a Poynting jet could be compressed by a factor a few and thus $L_e$ could in principle exceed $L$ by order unity.

\begin{figure}
\includegraphics[width=\columnwidth]{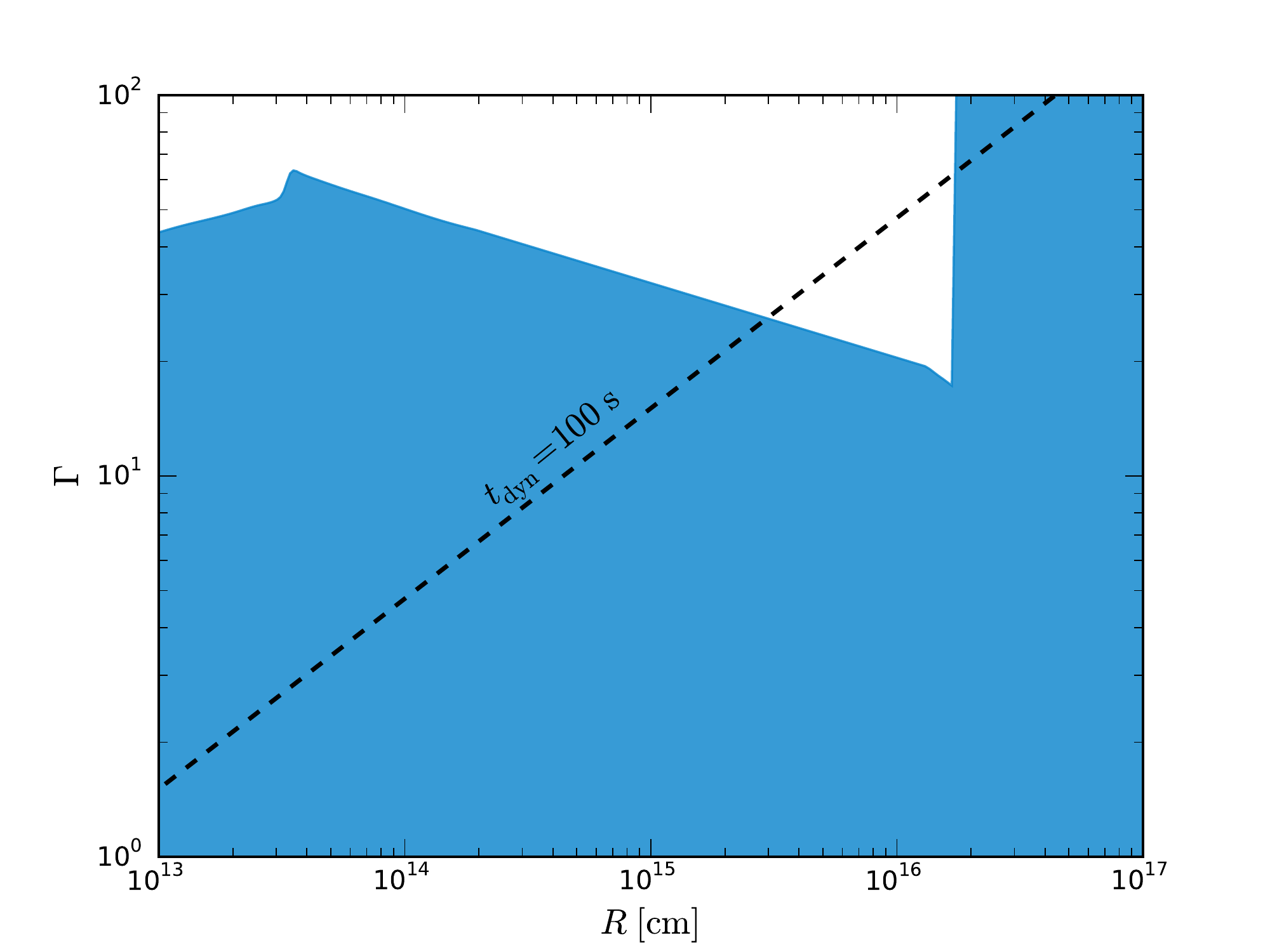}

\caption[The available parameter space for Magnetic Reconnection model of Sw J1644+57]{The available parameter space for any magnetic jet model that will not violate the LAT upper limits for Sw J1644+57. We assumed a 10\% efficiency for the jet, and matched the average keV flux of Sw J1644+57. The blue region is the parameter space where there won't be a LAT excess due to self-Compton scattering of keV photons. The sharp rise at $10^{14}$ cm is the where the electrons producing the hot X-rays ($>10$ keV) become Klein-Nishina suppressed. The sharp rise at $\sim10^{16}$ cm occurs because the electrons producing the soft X-rays become Klein-Nishina suppressed as well. Other considerations may rule out some of the space in blue; for instance it is unlikely that $\Gamma\gg 10$.}
\label{fig:R_Gam_Poynting}
\end{figure}

The Compton $\widetilde{Y}$, see eq (\ref{eq:ComptonY}), of the radiating electrons is
\begin{equation}
\widetilde{Y}=10\  \nu_\mathrm{keV}f_{\nu, \mathrm{mJy}} L_{48}^{-1}d_{L,28}^2
\end{equation}
Assuming an X-ray efficiency of 10\% (i.e., $L_{48}\sim 3$) and a flux at 1 keV of $f_{\nu,\mathrm{mJy}}\sim0.1$, we find that $\widetilde{Y}$ is $\sim 0.1$ This value means that if the jet is transparent to the peak of the SSC emission, there will be an LAT excess unless the SCC flux is suppressed due to Klein-Nishina effects. As $R$ increases, the opacity due to pair production decreases, but $\gamma_i$ also increases. At about $R\sim 10^{16} \mathrm{cm}$, the entire XRT band is Klein-Nishina suppressed, and the LAT constraint is no longer an issue as long as the minimum Lorentz factor of the electrons is $\sim \gamma_i$. We calculated the expected integrated flux SSC in the LAT band using eq (\ref{eq:flux_SSC}) accounting for both Klein-Nishina suppression and using the observed X-ray flux and an electron distribution given by $n=(\gamma/\gamma_i)^{-2.6}{N_e}/(4\pi R^2 (R/\Gamma))$, using the $\gamma_i$, $N_e$ given in eqs (\ref{LFi}) \& (\ref{Ne}) . If the integrated flux is greater than $8.5\times10^{-11}\ \mathrm{erg/s/cm^2}$, then the parameter space is ruled out for any Poynting dominated model with $\sim10\%$ efficiency. We show the allowed $R$---$\Gamma$ parameter space in Figure \ref{fig:R_Gam_Poynting}. 

Assuming that there are $\eta_p$ protons for every electron with LF $\sim\gamma_i$.\footnote{If there are no $e^\pm$ pairs, $\eta_p\geq 1$. It is unlikely that the jet will be pair-dominated, because for the majority of the kinetic energy at the base of the jet to be in pairs there must be 1000 pairs to every proton. The temperature at the base of the jet is simply not high enough for electron positron pairs to be thermal. If there are a significant number of pairs in the jet $\eta_p$ could be less than one and the estimated magnetization parameter would be larger than the value estimated in this section.} the luminosity carried by protons is
\begin{equation}
L_p=5\times10^{41}\frac{\Gamma^3\eta_p f_{\nu,\mathrm{mJy}}d_{L,28}^2}{L_{48}^{1/2}(1+z)}
\end{equation}
The magnetization parameter $\sigma\equiv B'^2/(8\pi n'_e m_p c^2)=L/L_p$ is then
\begin{equation}\label{eq:mag_param}
\sigma = 2\times10^6\ \Gamma^{-3}\eta_p^{-1}f_{\nu, \mathrm{mJy}}^{-1}L_{48}^{3/2}d_{L,28}^{-2}(1+z)
\end{equation}
Assuming there are 20 protons for every accelerated electron, $\Gamma\sim20$, the overall luminosity of the jet $L_{48}\sim 3$ (corresponding to a radiative efficiency of 10\%), and $f_{\nu, \mathrm{mJy}}\sim 0.1$, the magnetization parameter $\sigma\sim2800$ at the emission radius. At the base of the jet the magnetization is larger by a factor of $\Gamma$ and is $5.5\times10^{4}$. These values mean that the jet is highly magnetized, unless a small number of electrons are radiating or Sw J1644+57 is very efficient at converting its Poynting flux to radiation. If there are 100 protons per accelerated electron and the radiative efficiency is \(\sim10\%\), the magnetization (at the emission region) would be $\sim 280$ and $\sim 5600$ at the base of the jet (assuming a $\Gamma \sim 5590$).

Following \citet{KumarCrumley2015}, we assume that inside of a causally connected region of the jet, there are a number of acceleration sites where electron acceleration takes place. On average electrons, spend a time $\xi t'_\mathrm{dyn}$ in a current sheet, and they spend $\zeta t'_\mathrm{dyn}$ of their time outside of the reconnection regions. Inside of these regions, the electrons are heated to a power-law distribution. The hot electrons leave the acceleration regions and then cool through synchrotron radiation. The resulting observed synchrotron spectrum is a superposition of the radiation from electrons inside of the reconnection regions and electrons that have left the reconnection sites. The cooling electron Lorentz factor is
\begin{equation}
\gamma_c=\frac{7}{\zeta}\Gamma_1^3R_{16}L_{48}^{-1},
\end{equation}
with a corresponding synchrotron frequency of
\begin{equation}\label{eq:mag_nu_c}
\nu_c=3\times10^{-7}\ \mathrm{keV}\ \zeta^{-2}\Gamma_1^6R_{16}L_{48}^{-3/2}(1+z)^{-1}
\end{equation}
If $\nu_c \sim .1\ \mathrm{keV}$, i.e., when $R$, $\Gamma$, are large and $\zeta\ll 1$, the Poynting jet model will work as long as $\nu_i\sim0.1$ keV. If the X-ray emission is dominated by electrons inside of acceleration regions, the expected value for $p$ is 2.6, and if it is dominated by the electrons outside of the acceleration regions, $p= 1.6$. For a large swath of the parameter space, $\nu_c$ is small enough to potentially cause an infrared excess unless $R$ is small enough that the IR emission is self absorbed (essentially the same part of the parameter space that worked for the synchrotron model in \S \ref{sec:Synch}). However, as we will now show, the emission at 1 keV is likely dominated by electrons inside of the acceleration region. This means that it is no longer true that the spectrum between $\nu_i$ and $\nu_c$ is $\nu^{-1/2}$. If the emission at $\nu_i$ is greatly dominated by electrons inside of accelerations sites, the flux below $\nu_i$ will be $\propto\nu^{1/3}$ close to $\nu_i$ and it will soften to the $\propto\nu^{-1/2}$ at frequencies much less than $\nu_i$.

\begin{figure}
\includegraphics[width=\columnwidth]{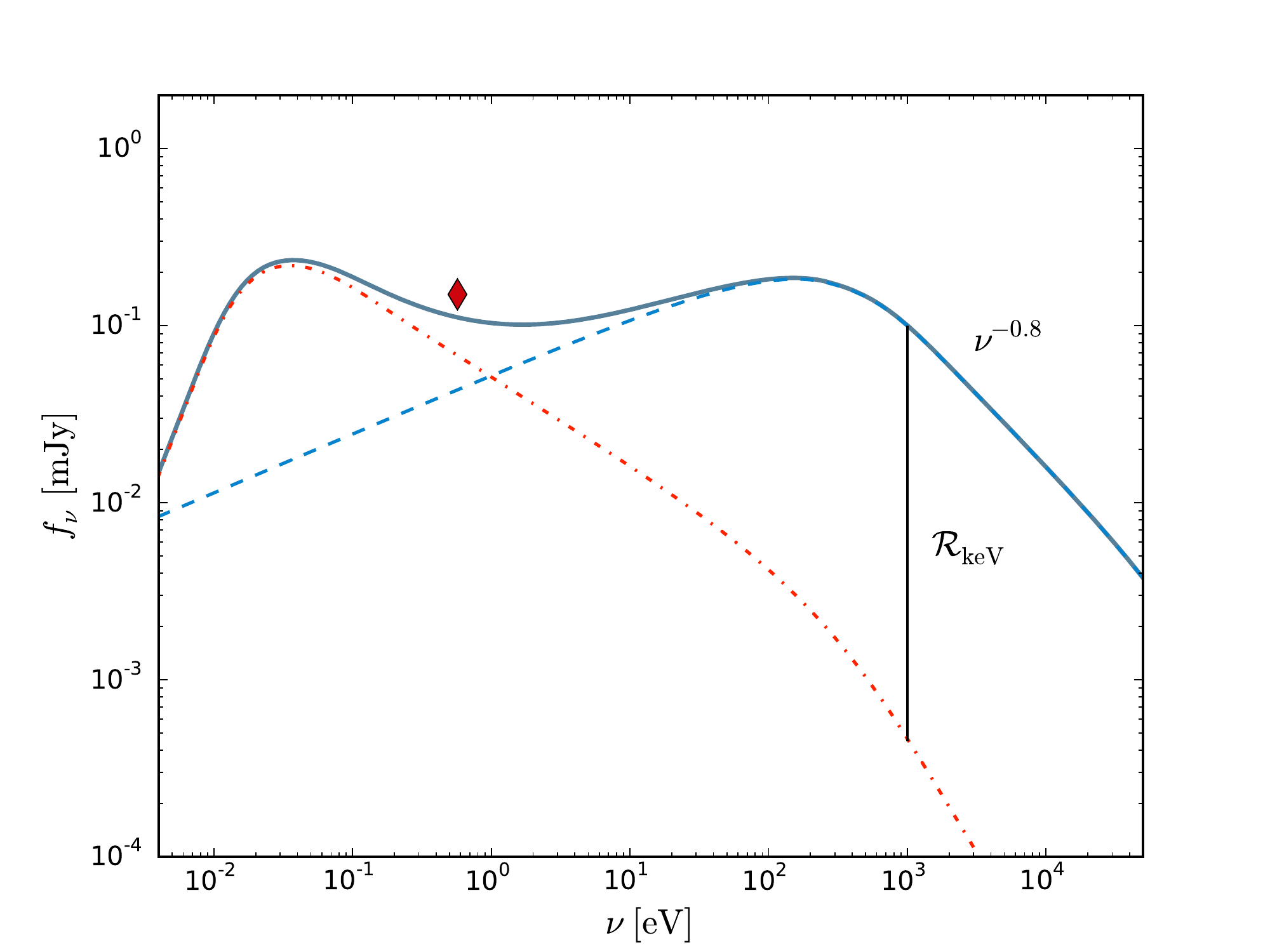}

\caption[A sample spectra in the magnetic reconnection process]{A sample spectrum from a Poynting dominated jet matching the observed spectrum of Sw J1644+57 is shown as a solid line. The contributions to the observed flux from the electrons in the particle acceleration region (blue dashed line) and the electrons outside of current sheets (red dash-dotted line). The ratio of the flux from electrons inside of particle acceleration regions to the flux from electrons outside is labeled $\mathcal{R}_{\rm keV}$. To avoid a K-band excess, $\mathcal{R}_{\mathrm{keV}}$ must be greater than 30, or $\nu_c$ must be greater than 100 eV. The K-band flux is the red diamond plotted at $\sim 1$ eV. The parameter were chosen as follows Luminosity in the jet is $3\times10^{48}$ erg/s, $\xi=5\times10^{-3}$, $\zeta = 1\times10^{-2}$, $R=5\times10^{14}\ \mathrm{cm}$, and $\Gamma=10$. }
\label{fig:Poynt_spectra}
\end{figure}

The ratio of synchrotron flux inside of the particle acceleration region to the flux outside of the particle acceleration region at 1 keV is calculated as in \citet{KumarCrumley2015}
\begin{equation}
\mathcal{R_\mathrm{keV}}\approx \frac{\xi t'_\mathrm{dyn}}{t'_\mathrm{cool}(\gamma_i)}.
\end{equation}
\begin{equation}\label{eq:ratio_keV}
\mathcal{R_\mathrm{keV}}=6\times10^{3}\xi L_{48}^{3/4}\nu_\mathrm{keV}^{1/2}(1+z)^{1/2}\Gamma_1^{-3}R_{15}^{-1/2}
\end{equation}
This ratio is $\gg 1$ for the allowed parameter space of Sw J1644+57 as long as $\xi$ is not very small (\(\lae 10^{-3}\)). Because \(\mathcal{R_\mathrm{keV}}\gg 1\), the observed flux is mostly coming from electrons inside of the acceleration region. When this is the case, an infrared excess will be avoided as long as $\mathcal{R}_\mathrm{keV}$ is sufficiently large. This is shown graphically in figure \ref{fig:Poynt_spectra}

The keV flux contribution from outside of particle acceleration regions is roughly $f_\mathrm{keV}/\mathcal{R}_\mathrm{keV}$. From eq (\ref{eq:mag_nu_c}) we know that $\nu_c$ is likely below the K-band, so the expected infrared flux is \(\sim \mathcal{R}_\mathrm{keV}^{-1} f_\mathrm{keV}\left(\nu_\mathrm{K-band}/\nu_\mathrm{keV}\right)^{-1/2}\). The flux in the K-band is roughly the same as the average flux at 1 keV, so to prevent particles outside of the particle acceleration region from overproducing the K-band flux, the ratio $\mathcal{R_\mathrm{keV}}$ must be greater than 30. $\mathcal{R_\mathrm{keV}}\gae 30$ can be achieved for the entirety of the radial parameter space plotted in Figure \ref{fig:R_Gam_Poynting} as long as $\Gamma_1$ is less than a few. An example of a spectrum that matches the observed properties is plotted in figure \ref{fig:Poynt_spectra}

One small downside with having the keV emission come from the reconnection regions is that the electron spectral index, $p$, is related to the observed photon spectral index, $\beta$ as \(\beta = (p-1)/2\). So for a $\beta$ of 0.8, $p$ should be approximately 2.6. This value is softer than what is observed in particle-in-cell simulations of magnetic reconnection, that generally see a $p<2$ electron distribution. The softer spectrum could be attributed to the fact that, at any time, we are seeing emission from multiple particle acceleration regions. At late times ($t>10^7$ s), the spectrum hardens from $\beta=0.8$ to $\beta =0.4$ \citep{Saxton2012}. During this time, $p\sim 1.8$, which is consistent with electron PIC simulations.

\section{Conclusions}
A tidal disruption event with a relativistic jet like Sw J1644+57 presents a unique opportunity to test the non-thermal emission models used in modeling other relativistic jets such as the ones observed in AGN and GRB. Sw J1644+57 had a period of intense, high-luminosity flaring that lasted $\sim 10$ days, allowing for plenty of time to do multi-wavelength measurements in the $\gamma$-rays, optical, infrared, and radio. These measurements put strong constraints on the allowed parameter space of any emission model. The Thomson photosphere for TDE jets is close to the launching site, i.e., the Schwarzschild radius, which makes the analysis of radiative processes much simpler than say GRB jets (which have optical depth $>10^6$ at the launching site) The TDE creates the accretion disk that feeds the super-massive black hole, the total energy budget of Sw J1644 is well constrained at $\sim 1 M_{\odot}c^2$, and this also helps narrow down the radiation mechanism at work in TDEs.

In this paper, we comprehensively considered emission mechanisms for the Swift XRT-BAT radiation observed in the tidal disruption event Sw J1644+57. We have explored a wide variety of different radiation mechanisms for the X-ray photons: synchrotron (when particles are accelerated in shocks and in reconnection zones), internal and external IC, and Comptonization below the Thomson photosphere. We find that several of these mechanisms are consistent with multi-wavelength data for J1644+57 as long as the jet is magnetic dominated.  The strengths and weaknesses of the various mechanisms are summarized below. 

We robustly rule out internal inverse Compton models for the Sw J1644+57, because it leads to too much flux in the infrared band; electrons cool due to IC scatterings and that causes the X-ray spectrum to extend to IR whereas observations show that the spectrum hardens not far below 0.1 keV. 

We developed a new way to calculate the synchrotron radiation from shocks that accounts for Klein-Nishina effects, self-consistently calculates the synchrotron self-absorption frequency when $\nu_c<\nu_a$, and is applicable to non power-law electron distributions. We find the synchrotron model with a single episode of acceleration of particles has difficulties satisfying all observational constraints. There does exists a small area at small radii ($R\lae 2\times10^{13}\ \mathrm{cm}$), deep below the photosphere, where synchrotron self-absorption can prevent the K-band from being overproduced by cooled electrons (See Fig \ref{fig:Synch}). Since the emission is in the fast cooling regime, the required electron index, $p$, is quite hard $\sim 1.6$ at early times and $\sim 0.8$ at late times; $\mathrm{d}n_e/\mathrm{d}E\propto E^{-p}$. Since such a hard electron index is difficult to achieve in shocks and the allowed parameter space is beneath the photosphere, synchrotron radiation produced in shocks is an unlikely source of the X-rays of Sw J1644+57.

We found that proton synchrotron model for the X-ray emission of Sw J1644+57 may be able to explain the observed spectrum, if the jet is Poynting flux dominated. The proton synchrotron model requires a large minimum proton Lorentz factor $\gae 10^4$, and requires a magnetization parameter $\sigma\gae 10^{4}$. The proton model has relatively low radiative efficiency of $\sim 1\%$, and so it is an unlikely source of the radiation of Sw 1644+57.

We also considered external inverse Compton of UV radiation from the accretion disk wind by the relativistic jet. We find that the EIC model is marginally consistent with the observations if the scattering happens above the photosphere of the wind, the efficiency in launching the wind is large, and the Lorentz factor of the jet is not too large, $\Gamma\lae5$. The EIC model may need to be finely tuned to match the X-ray light curve. The X-ray luminosity of Sw J1644+57 roughly matches the theoretical TDE fallback rate. The agreement between the fallback rate and the X-ray light curve strongly suggests that the X-ray luminosity is tracking the jet power, and that the jet converts a constant fraction of its energy into electrons. These hot electrons then quickly IC scatter the external radiation field into the X-rays. As we showed in section \ref{sec:EIC}, requiring that the observed X-rays track the total jet power places an additional constraint on the EIC mechanism: the radius of emission must be proportional to the luminosity of the external radiation field (i.e., $L_{\rm ERF}/R\sim$ constant). The quantity $L_{\rm ERF}/R$ is highly uncertain because the large extinction of the host galaxy of Sw J1644+57 caused a lack of UV and optical data for this TDE. However, if $L_{\rm ERF}/R$ fluctuates by a factor more than a few, the EIC model requires fine tuning.

Photospheric models, where a thermal seed spectrum is reprocessed through inverse Compton scatterings by hot electrons, face severe difficulties in reproducing the observed spectrum of Sw J1644+57. The main problems is due to the fact that the hot electrons that are scattering photons lose too much of their energy before they are able to populate a hard spectrum, such as the one observed in Sw J1644+57. It may be possible to match the spectrum at early times if there are many ($\gae100$) reheating events just below the photosphere in Sw J1644+57, or for very inefficient jets with $<1\% $ efficiency. However, at late times when the luminosity of the jet is much lower, the photosphere of a relativistic jet is inside of the Schwarzschild radius of a $\sim 10^6 M_\odot$ black hole, and the photospheric proccess is no longer a viable method of producing the X-rays in Sw J1644+57.

Finally we considered magnetic reconnection in a Poynting-dominated jet. We find that the luminosity and the spectrum can be easily matched with a model of magnetic reconnection based off of very general principles for a large swath of the available parameter space. The magnetization of the jet is large $\sigma \sim 250-2800$. The emission is dominated by electrons inside of reconnection regions where the electrons are unable to cool. Outside of the reconnection region, they cool very quickly through synchrotron radiation. But because the number of electrons that radiate at 1 keV is much higher inside of the reconnection region, the cooled electrons will not cause an infrared excess. We feel that a Poynting-dominated jet with magnetic reconnection is the most likely model for the X-rays observed in Sw J1644+57.

While we are unable to uniquely identify one radiation mechanism that alone can explain the observed X-ray flux of Sw J1644+57, we do find a set of common requirement for all successful models which are: electrons must be prevented from cooling in order to avoid a K-band excess, the electrons must be accelerated into a hard power-law of electrons with $p\approx1.6$, or the magnetic luminosity must be large $L_B>L_X$. Since all of these characteristics are typical of magnetic reconnection in a Poynting jet,  there is strong evidence that Sw J1644+57 launched a Poynting jet.

\section*{Acknowledgements}
The authors would like to thank Volker Bromm and Milos Milosavljevic for generously providing computational resources. PC acknowledges financial support from the WARP program of the Netherlands Organisation for Scientific Research (NWO). This work made use of data supplied by the UK Swift Science Data Centre at the University of Leicester.

\bibliographystyle{mnras}
\bibliography{SwJ1644.bib}

\appendix
\section{Synchrotron Self-Compton Model}\label{sec:SSC}
Here we describe the new numerical methods used to self-consistently calculate the the SSC flux from an electron distribution with a co-moving number density  $dn' = n'_e d\gamma$. Because we are worried about calculating the SSC spectrum for self-absorbed spectra, we must develop a new methodology because previous methods such as \citet{Nakar2009} may not calculate the inverse Compton power correctly when the seed spectrum is harder than $f_\nu \propto\nu$ is Klein-Nishina suppressed. In this appendix, as before, primed quantities are in the co-moving rest frame, and unprimed quantities are in the observer rest frame. Unlike the main paper, all quantities in the appendix are in CGS units (including $\nu$ and $f_\nu$).
\subsection{Synchrotron Radiation}
The synchrotron power radiated by a single electron with a Lorentz factor $\gamma$ peaks at frequency in the co-moving rest frame $\nu'_p=qB' \gamma^2/(2\pi m_e c)$. $P_{\nu'}$, the angle-averaged power radiated at a given frequency $\nu'$ , depends on ratio $x\equiv \nu'/\nu'_p$ and the magnetic field $B'$. 
\begin{equation}\label{eq:synch_pow_nu}
P_{\nu'}\approx\frac{\sqrt{3}q^3B'}{m_e c^2} F\left(\frac{\nu'}{\nu'_p}\right)
\end{equation}
The exact formula for $F(x)\equiv F(\nu'/\nu_p')$ in terms of integrals over modified Bessel functions can be found in \citet{Ginzburg1965}. However, to calculate the Synchrotron power radiated, we patch together their third-order approximation to $F(x)$ when $x\ll1$ and first-order approximation of $F(x)$ when $x\gg 1$. The approximation agrees well with the exact solution to the synchrotron power, underestimating the true power slightly at $x\approx 1$ and overestimating synchrotron power slightly when $x\gg 1$, but the approximation has the benefit that it is integrable. We must integrate over this function later to calculate the total synchrotron power radiated by a single electron.
\begin{equation}
F(x)=
\left\{\begin{array}{ll}\hskip -3pt
c_0 x^{1/3}
g(x) & \mbox{for } x\leq 1\\ \\
\hskip -3pt
c_0 g(1)
\sqrt{x}\exp{(1-x)} & \mbox{for } x>1
\end{array}\right.
\end{equation}
where
\begin{equation}
g(x)=
1-
\frac{\Gamma(\frac{1}{3})}{2}\left(\frac{x}{2}\right)^{2/3} 
+ \frac{3}{4}\left(\frac{x}{2}\right)^2 - 
\frac{9}{40}\frac{\Gamma(\frac{1}{3})}{\Gamma(\frac{5}{3})}
\left(\frac{x}{2}\right)^{10/3}
\end{equation}
\begin{equation}
c_0 = \frac{4\pi}{\sqrt{3}\Gamma(\frac{1}{3})}2^{-1/3}\approx 2.15
\end{equation}
and $\Gamma(x)$ is the Gamma function. 

The co-moving synchrotron specific intensity, $I_\nu'$, is calculated using the formal solution to the 1-D radiative transfer equation of a constant synchrotron source with no incident radiation through an optical depth $\tau_{\nu'}$ ignoring scattering, \citet[cf.][]{RybickiLightman}
\begin{equation}\label{eq:rad_transfer}
I_{\nu'} = (1 - e^{-\tau_{\nu'}})\frac{j_{\nu'}}{\alpha_{\nu'}}
\end{equation}
The absorption coefficient, $\alpha_{\nu'}$, for electron distribution emitting a synchrotron power $P_{\nu'}$ is 
\begin{equation}
\alpha_{\nu'} = -\frac{c^2}{8\pi\nu'}\int{d\gamma\ \gamma^2 P_{\nu'}
\frac{\partial}{\partial \gamma}\left[\frac{n'_e}{\gamma^2}\right]}.
\end{equation}
The emission coefficient, $j_{\nu'}$ is
\begin{equation}
j_{\nu'}=\frac{1}{4\pi}\int{d\gamma n'_e P_{\nu'}}.
\end{equation}
We calculate the optical depth to synchrotron absorption by assuming the source has a thickness $R/2\Gamma$:
\begin{equation}\label{eq:syn_opt_depth}
\tau_{\nu'} \approx \alpha_{\nu'} \frac{R}{2\Gamma}.
\end{equation}
We also use the synchrotron optical depth to calculate the self-absorption frequency, $\nu'_a$. $\nu'_a$ is found by numerically solving the equation $\tau_{\nu'_a} =1$.

Assuming that the source is spherically symmetric and isotropic and moving at a bulk Lorentz factor $\Gamma\gg 1$, the observed synchrotron flux at frequency $\nu$, $f_{\nu, \mathrm{syn}}$, is related to $I_{\nu'}$ by
\begin{equation}\label{eq:flux_syn}
f_{\nu, \mathrm{syn}} \approx 4\pi\Gamma(1+z) I_{\nu'}\left(\frac{R}{d_L}\right)^2
\end{equation}

Note that ignoring scattering effects in eq (\ref{eq:rad_transfer}) means that the previous equation for the flux is only valid when the optical depth to Thomson scattering is small, \textit{ i.e.} \(\sigma_T R\int{d\gamma n'_e/(2\Gamma)}\ll 1\)

The total synchrotron energy radiated by an electron with Lorentz factor $\gamma$ in the co-moving frame can be calculated by integrating over $P_{\nu'}$. However, any photons radiated at frequencies less the self-absorption frequency will be quickly absorbed by other electrons. These absorbed photons act as a heating source \citep{Ghisellini88}. We approximate this synchrotron heating by only allowing photons with frequencies $\geq\nu'_a$ to carry away energy from the electrons

\begin{equation}
P_\mathrm{syn}(\gamma) = \int_{\nu'_a}^\infty {d\nu' P_{\nu'}}
\end{equation}
Defining a new quantity $x_a\equiv \nu'_a/\nu'_p$ and integrating over eq (\ref{eq:synch_pow_nu}) for $P_{\nu'}$, we find the following expression for the synchrotron power emitted by an electron with Lorentz factor $\gamma$
\begin{equation}\label{eq:correct_synch}
P_\mathrm{syn}(\gamma)\approx\frac{\sqrt{3}q^3B'}{m_e c^2} G\left(x_a\right).
\end{equation}

\begin{equation}
G(x_a)=
\left\{\begin{array}{ll}\hskip -3pt
c_0\left[h(1)- x_a^{4/3}
h(x_a)\right] + c_1 & \mbox{for } x_a\leq 1\\ \\
\hskip -3pt
c_0 g(1)\left[\sqrt{x_a}
\exp{(1-x_a)} \right.\\
\left.\quad\quad\ \ +\frac{e}{2}\sqrt{\pi}\mathrm{erfc}(\sqrt{x_a}) \right]& \mbox{for } x_a>1
\end{array}\right.
\end{equation}
where erfc is the complementary error function, 
\begin{equation}
h(x) = \frac{3}{2} 
-\frac{\Gamma(\frac{1}{3})}{2}\left(\frac{x}{2}\right)^{2/3} 
+ \frac{9}{20}\left(\frac{x}{2}\right)^2 - 
\frac{27}{280}\frac{\Gamma(\frac{1}{3})}{\Gamma(\frac{5}{3})}
\left(\frac{x}{2}\right)^{10/3}
\end{equation}
and
\begin{equation}
c_1= c_0g(1)\frac{\sqrt{\pi}}{2}\mathrm{erfc}(1)\approx 0.822.
\end{equation}
When $x_a\ll 1$ eq (\ref{eq:correct_synch}) agrees with the standard synchrotron power formula $\frac{4}{3}\sigma_t c \gamma^2B^2/(8\pi)$ within 1\%.

\subsection{Synchrotron Self Compton}
We use the synchrotron flux $f_{\nu, \mathrm{syn}}$, eq (\ref{eq:flux_syn}), as the seed photons to be inverse Compton scattered to calculate the SSC flux. In doing so, we implicitly assume that a photon is only inverse Compton scattered once, which is a good assumption when either $Y\ll 1$ or the second scattering is Klein-Nishina suppressed. Our code can be easily extended to handle multiple scatterings self-consistently, but it becomes too numerically intensive to calculate multiple scatterings quickly.

\begin{figure}
\includegraphics[width = \columnwidth]{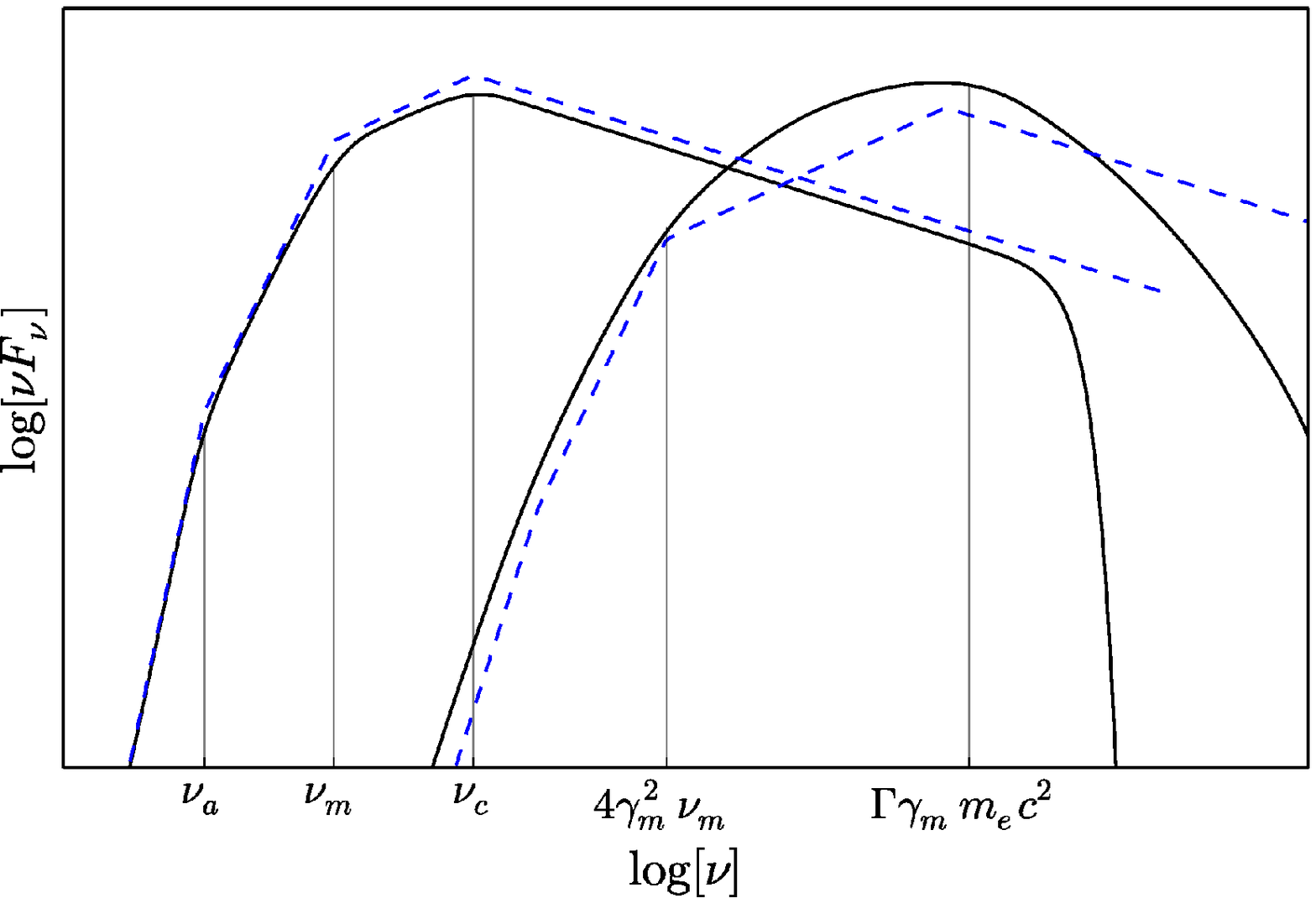}
\caption[Example Synchrotron Self-Compton Spectra]{Example SSC spectra. The inverse Compton is exact to corrections of $\mathcal{O}(\gamma^{-2})$ to arbitrarily large $\nu$. In Figure \ref{fig:Synch}, the cooling is calculated self-consistently, but to make our spectrum easier to compare with previous results in this figure, the cooling is calculated assuming synchrotron cooling only. The curve that peaks at $\nu_i$ is the synchrotron component, and the curve that peaks at $\Gamma\gamma_i m_e c^2$ is the inverse Compton component. The dashed power-law synchrotron approximations are given in \citet{Sari1998}, and inverse Compton dashed lines are calculated assuming the scattering takes place in the Thomson regime using power-law approximations from \citet{Sari2001}. The spectra agree well below the frequency $\Gamma\gamma_m m_e c^2$. Above this frequency, we should expect deviations due to Klein-Nishina effects. Above the Klein-Nishina frequency, the spectrum is similar to the spectrum found in \citet{Nakar2009,Wang2010}.}
\label{fig:SSC_spectra}
\end{figure}
We calculate the SSC flux using the equation for the inverse Compton spectrum produced by a single electron traveling through an isotropic field of photons given in \citet{Jones1968,Blumenthal1970}. This equation differs from the exact Klein-Nishina cross section only by terms $\mathcal{O}(\gamma^{-2})$. This equation is used to calculate the inverse Compton flux at arbitrarily large $\nu_\mathrm{obs}$ as it accounts for the Klein-Nishina corrections to the Thomson cross-section. Using the observed synchrotron flux calculated with eq (\ref{eq:flux_syn}) to calculate the isotropic photon field and integrating over the co-moving electron density distribution yields the following equation for the SSC flux at $\nu_\mathrm{obs}$
\begin{equation}\label{eq:flux_SSC}
f_{\nu_\mathrm{obs, IC}} =
\frac{3\nu_\mathrm{obs}}{8\Gamma}\sigma_T R\int{ d\gamma \frac{n'_e}{\gamma^2}\int_{\nu_\mathrm{min}}^{\nu_\mathrm{obs}}{d\nu \frac{f_{\nu,\mathrm{syn}}}{\nu^2} g(\epsilon_\nu, q)}}
\end{equation}
The function $g(\epsilon_\nu, q)$ is a function that accounts for the energy and angular dependence of the Klein-Nishina cross section. $g(\epsilon_\nu, q)$ depends on two dimensionless parameters: $\epsilon_\nu$, four times the photon's energy in the co-moving frame normalized by the electron energy, and $q$.
\begin{equation}\label{eq:KN_angle}
g(\epsilon_\nu, q) = 1+q-2q^2+2q \ln{q} +\frac{(\gamma^2 \epsilon_\nu q)^2} {2+2\gamma^2 \epsilon_\nu q}(1-q)
\end{equation}
\begin{equation}
\epsilon_\nu =\frac{4h\nu(1+z)}{\Gamma \gamma m_e c^2};\quad\quad
q=\frac{1}{\gamma^2 \epsilon_\nu}\frac{\epsilon_{\nu_\mathrm{obs}}}{(4-\epsilon_{\nu_\mathrm{obs}})}
\end{equation}
The Thomson regime corresponds to $\epsilon_\nu\ll 1/\gamma^2$.

The lower limit of the $d\nu$ integral is found by numerically solving the following equation for $\nu_\mathrm{min}$:
\begin{equation}\label{eq:nu_min_exact}
\epsilon_{\nu_\mathrm{obs}}=\frac{4\gamma^2\epsilon_{\nu_\mathrm{min}}}{1+\gamma^2\epsilon_{\nu_\mathrm{min}}}
\end{equation}
When $\nu_\mathrm{obs}$ is well within the Thomson regime, \textit{ i.e.} $h\nu_\mathrm{obs}\ll \Gamma \gamma m_e c^2/(1+z)$, the last term in eq (\ref{eq:KN_angle}) can be ignored, and the eq (\ref{eq:nu_min_exact}) can be linearized and solved for $\nu_\mathrm{min}$:
\begin{equation}\label{eq:nu_min_thom}
\nu_\mathrm{min}\approx\frac{\nu_\mathrm{obs}}{4\gamma^2} \quad\mathrm{when} \quad \epsilon_{\nu_\mathrm{obs}} \ll 1
\end{equation}
In the Thomson regime, $\nu_\mathrm{min}$ is so small it is usually taken to be zero, but in the Klein-Nishina regime $\nu_\mathrm{min}$ will approach $\nu_\mathrm{obs}$. $\nu_\mathrm{min}$ will be larger than $\nu_\mathrm{obs}$ when $h\nu_\mathrm{obs}(1+z)\approx\Gamma\gamma m_e c^2$. At this point, the $d\nu$ integral is zero.

Our eq (\ref{eq:flux_SSC}) simplifies to the expression often seen in in previous works concerned with SSC flux in the Thomson regime, such as \citet{Sari2001} eq (A1), after making $\nu_\mathrm{min}= 0$, assuming $\epsilon_{\nu_{\mathrm{obs}}}\ll 4$ to make $q$ first order in $\epsilon_{\nu_\mathrm{obs}}$, removing the last term from eq (\ref{eq:KN_angle}) for $g(\epsilon_{\nu}, q)$, and a simple variable substitution in the $d\nu$ integral. However, our eq (\ref{eq:flux_SSC}) for $f_{\nu_\mathrm{obs, IC}}$ is accurate to arbitrarily large $\nu_\mathrm{obs}$ when using eq (\ref{eq:nu_min_exact}) to find $\nu_\mathrm{min}$ and the full form of $g(\epsilon_{\nu}, q)$. Since we made no assumptions on the functional forms of $n'_e$, other than $\gamma\gg1$, our formula may be used to calculate the SSC flux for any given electron distribution $n'_e$ that is integrable and differentiable. A sample SSC spectrum is shown in Figure \ref{fig:SSC_spectra}.

Since the inverse Compton flux will be calculated at large photon energies, $\gae 100$ MeV, we must take into account the $\gamma+\gamma$ pair opacity. We use the methodology of \citet{Lithwick2001} to calculate the optical depth of pair production $\tau_{\gamma\gamma}$. If $\tau_{\gamma\gamma}\geq1$ for a particular frequency, we assume those photons are absorbed by pair-production and multiply the observed flux at that frequency by a factor $\exp{\{-\tau_{\gamma\gamma}\}}$.
\begin{equation}\label{eq:tau_gammagamma}
\tau_{\gamma\gamma}(\nu_\mathrm{obs})
=\frac{11}{180}\frac{\sigma_T d_l^2}{\Gamma^2 R h c(1+z)}\int_{\nu_{\gamma\gamma}}^\infty{d\nu\frac{f_\nu}{\nu}}
\end{equation}
\begin{equation}\label{eq:gammagamma_min}
\nu_{\gamma\gamma}=\left(\frac{\Gamma m_e c^2}{h(1+z)}\right)^2\frac{1}{\nu_\mathrm{obs}}
\end{equation}

To calculate the total inverse Compton cooling, we integrate the inverse Compton spectrum of a single electron like we did for synchrotron radiation. This integral has been already calculated exactly for arbitrary seed photons and electron Lorentz factor $\gamma$ by \citet{Jones1965}. We apply his solution to SSC cooling, adding the additional constraint that $\gamma\gg1$
\begin{equation}\label{eq:pow_IC}
P_\mathrm{IC}(\gamma)= \frac{12\sigma_T}{\left(\Gamma\gamma^2
\right)^2}\left(\frac{d_L}{R}\right)^2\int_{0}^\infty{d\nu f_{\nu,\mathrm{syn}} F_\mathrm{IC}(\epsilon_\nu,\gamma)}
\end{equation}

\begin{multline}
F_\mathrm{IC}(\epsilon_{\nu},\gamma) =
\left[f_1(\gamma^2\epsilon_\nu/2)-f_1(\epsilon_\nu/8)\right]/\epsilon_\nu^3\\
{ -\left[f_2(\gamma^2\epsilon_\nu/2)-f_2(\epsilon_\nu/8)\right]/(4\epsilon_\nu^2)}
\end{multline}

\begin{align}
f_1(x) &= (x+6+3/x)\ln{\left(1+2x\right)} \nonumber \\
 &  \qquad-(22x^3/3+24x^2+18x+4)(1+2x)^{-2}\nonumber \\
 &  \qquad\qquad -2+2\mathrm{Li_2}\left(-2x\right)\\
  &\approx 8x^3/9 \quad \mathrm{when\ }x\ll 1
\end{align}
\begin{align}
f_2(x) &= [x+31/6+5/x+3/(2x^2)]\ln{\left(1+2x\right)} \nonumber\\
 &\qquad-(22x^3/3+28x^2+103x/3+17+3/x)\nonumber\\
  &\qquad\qquad \times (1+2x)^{-2}-2+\mathrm{Li_2}\left(-2x\right)\\
  &\approx 4x^2/3 \quad \mathrm{when\ }x\ll 1
\end{align}
The function $\mathrm{Li}_2$ is called the dilogarithm, and it is defined as
\begin{equation}
\mathrm{Li_2}(x)=-\int_0^x{\frac{\ln{(1-x')}}{x'}dx'}.
\end{equation}
This equation for $P_\mathrm{IC}$ is exact (ignoring multiple scatterings of a single photon) for all $\gamma\gg 1$, working both in the Thomson and Klein-Nishina regime. While it is a complicated formula, it can be evaluated in a straightforward manner numerically. Most scientific coding languages have a predefined function for $\mathrm{Li_2}$, so calculating the inverse Compton power is only a single numerical integration over the seed spectrum. 

The total SSC power in the co-moving rest frame is simply

\begin{equation}\label{eq:pow_SSC}
P_\mathrm{TOT}(\gamma)=P_\mathrm{syn}(\gamma)+P_\mathrm{IC}(\gamma)
\end{equation}

\subsection{The electron distribution}
We find the co-moving electron distribution by solving the continuity equation. Assume there is a source of electron $S(\gamma)$,
\begin{equation}
S(\gamma)=\left\{\begin{array}{ll}
\hskip -3pt 0 & \mbox{when } \gamma < \gamma_m \\
\hskip -3pt\displaystyle \dot{n}' \left(\gamma/\gamma_m\right)^{-p} &\mbox{when } \gamma\geq\gamma_m.
\end{array}
\right.
\end{equation}
Then the electron distribution is found by solving
\begin{equation}\label{eq:ContinuityEQ}
\frac{\partial n'_e}{\partial t}+
\frac{\partial}{\partial \gamma}\left(\dot{\gamma}n'_e\right)=
S(\gamma)
\end{equation}
We solve this equation approximately by breaking it into two parts: one where the electrons are effectively uncooled and one where cooling is important, determined by comparing the following two timescales:
\begin{equation}
t'_\mathrm{dyn}=\frac{R}{2c\Gamma},\quad\quad 
t'_\mathrm{cool}=\frac{m_e c^2\gamma}{P_\mathrm{TOT}(\gamma)}
\end{equation}
Defining a cooling electron Lorentz factor, $\gamma_c$, as the electron LF $\gamma$ where $t'_\mathrm{dyn} = t'_\mathrm{cool}$, the solution of the continuity equation is:
\begin{equation}\label{eq:electron_dist}
n'_e = 
\left\{\begin{array}{ll}\displaystyle
\hskip -3pt t'_\mathrm{dyn}S(\gamma) & \mbox{when } \gamma \leq \gamma_c \\
\hskip -3pt  \displaystyle \frac{m_e c^2}{P_\mathrm{TOT}(\gamma)}\int_{\gamma}^\infty{d\gamma_e S(\gamma_e)} & \mbox{when } \gamma > \gamma_c
 .
\end{array}
\right.
\end{equation}

We desire an $n'_e$ so that the synchrotron flux from $n'_e$ matches the observed X-ray flux $f_{\nu_i}$ at $\nu_i$ with a spectral slope $\beta$, ($f_\nu\propto\nu^{-\beta}$). We choose the source of the electrons, $S(\gamma)=\dot{n}'(\gamma/\gamma_m)^{-p}$, to ensure the X-ray flux is matched as follows: 

First we find the electron Lorentz factor whose synchrotron emission peaks at $\nu_i$  by inverting eq (\ref{eq:nu_i})
\begin{equation}\label{eq:gamma_i_comp}
\gamma_i = 2.8\times10^{5}
\sqrt{\frac{\nu_{i,\mathrm{keV}}(1+z)}{B\Gamma}}
\end{equation}
Then we match $\beta$ by requiring the injected electron index $p$ is  
\begin{equation}\label{eq:electron_spectrum}
p = 
\left\{\begin{array}{ll}\displaystyle
\hskip -3pt 2\beta + 1 & \mbox{if } \gamma_i < \gamma_c \\
\hskip -3pt  \displaystyle 2\beta+2-\left.\frac{d\log{P_\mathrm{TOT}}}{d\log{\gamma}}\right|_{\gamma=\gamma_i} & \mbox{if } \gamma_i \geq \gamma_c
 .
\end{array}
\right.
\end{equation}
where $P_\mathrm{TOT}$ is given in eq (\ref{eq:pow_SSC}). When the cooling is not dominated by Klein-Nishina cooling, $P_\mathrm{TOT}\propto\gamma^2$, and the standard cooling regime result  $\beta = p/2$ is recovered. 

Finally, we match the specific flux at $\nu_i$ by choosing a $\dot{n}'$ so that the synchrotron flux given in eq (\ref{eq:flux_syn}) is equal to $f_{i}$

Now we have described a complete set of equations that can be solved self-consistently for $f_\nu$ and $n'_e$ after specifying the following free parameters, $R$, $\Gamma$, $B'$, that will match an observed flux $f_i$ and spectrum $\beta$. These equations will work for any set of free parameters, but since we have assumed $\Gamma,\ \gamma\gg1$ in our derivation of the equations, we do not consider any electrons with $\gamma<2$ and require $\Gamma \geq 2$. The radiative transfer equation eq (\ref{eq:rad_transfer}) is no longer valid when the optical depth to Thomson scattering is greater than 1. 
\subsection{Energetics}
The isotropic equivalent luminosity carried by a particular component of the jet is calculated using the co-moving energy density $U'$
\begin{equation}
\mathcal{L}= 4\pi U' R^2c\Gamma^2
\end{equation}
The luminosity in the magnetic field is 
\begin{equation}\label{eq:MagLum}
\mathcal{L}_B = \frac{1}{2}B'^2 R^2 c\Gamma^2
\end{equation}
Solving eq (\ref{eq:gamma_i_comp}) for $B'$ we find
\begin{equation}
\mathcal{L}_B\approx 10^{60}\ \nu_{i,\mathrm{keV}}^2\gamma_i^{-4}R_{14}^2(1+z)^2\ \mathrm{erg/s}.
\end{equation}
Requiring the magnetic luminosity to not exceed $10^{50}$ erg/s places the following restriction on $\gamma_i$:
\begin{equation}
\label{eq:gamma_i_lim}
\gamma_i \geq 320\ \nu_{i,\mathrm{keV}}^{1/2}R_{14}^{1/2}(1+z)^{1/2}.
\end{equation}

The luminosity required to power the electrons producing the X-ray radiation is
\begin{align}
\label{eq:e_lum_tot}
\mathcal{L}_e & =  4\pi m_e c^3R^2\Gamma^2 t'_\mathrm{dyn}\int
{\gamma S(\gamma) \ d\gamma} \\
\mathcal{L}_e & \approx  \Gamma^2 \gamma_i m_e c^2 N_e\times\max{\left\{ \frac{1}{t'_\mathrm{dyn}},\frac{1}{t'_\mathrm{cool}}\right\}}.
\end{align}
In the vast majority of the parameter space, the electrons are in the fast cooling regime, and the inverse Compton cooling is less important than the synchrotron cooling. When synchrotron cooling is the dominant cooling mechanism, the luminosity required is
\begin{equation}
\mathcal{L}_e \approx 5\times10^{48}\ f_{i, \mathrm{mJy}} \nu_{i,\mathrm{keV}} d_{L,28}^2\ \mathrm{erg/s},
\end{equation}
when \(t'_\mathrm{dyn},t'_{c, \mathrm{IC}} \gg t'_{c, \mathrm{syn}}\), $\mathcal{L}_e\sim 3\times10^{47}\ \mathrm{erg/s}$ if $f_{i}=0.1\ \mathrm{mJy}$ and $d_{L,28}=0.58$. Therefore we arrive at the intuitive result that in the fast cooling regime, all of the energy given to the electrons by the shock will come out as radiation. If there are $\eta_p$ cold protons per electron, the luminosity carried by the protons is
\begin{equation}\label{eq:pro_lum_tot}
\mathcal{L}_p = 4\pi \eta_p m_p c^3 R^2\Gamma^2t'_\mathrm{dyn} \int
{S(\gamma)\ d\gamma}\approx \eta_p\frac{m_p}{m_e}\gamma_i^{-1}\mathcal{L}_e.
\end{equation} 
When  \(t'_\mathrm{dyn},t'_{c, \mathrm{IC}} \gg t'_{c, \mathrm{syn}}\), the luminosity carried by cold protons is 
\begin{equation}
\label{eq:Proton_lum}
\mathcal{L}_p\approx 9\times10^{51}\ \eta_p\gamma_i^{-1} f_{i, \mathrm{mJy}} \nu_{i,\mathrm{keV}} d_{L,28}^2\ \mathrm{erg/s},
\end{equation} or $\sim 3\times10^{50}\ \eta_p/\gamma_i \ \mathrm{erg/s}$ for the fiducial parameters. From the earlier constraint on $\gamma_i$, eq (\ref{eq:gamma_i_lim}), we see that the energy carried by the protons will not exceed $10^{50} \mathrm{erg}$ unless $\eta_p$ is greater than 100--1000.

We estimate location of the photosphere using equation (\ref{eq:e_lum_tot}) for the total electron luminosity.
\begin{equation}
R^2_{ph} = \frac{\mathcal{L}_e\sigma_T t'_{\rm dyn}}{\Gamma^2\gamma_i m_e c^2}
\end{equation}
The photosphere's location is shown in figure \ref{fig:Synch}.

\end{document}